\documentclass[journal]{new-aiaa}
\usepackage[utf8]{inputenc}
\usepackage{textcomp}

\usepackage{graphicx}
\usepackage{amsmath}
\usepackage{longtable,tabularx}
\usepackage{booktabs}
\usepackage{multirow}
\usepackage{subfig}
\usepackage{algorithm}
\usepackage{float}
\usepackage{algpseudocode}

\usepackage{lineno}

\graphicspath{{./}{JATM_manuscript/}{JAIS_manuscript/}}

\newcommand{\taumin}{0.5}

\algrenewcommand\algorithmicrequire{\textbf{Input:}}
\algrenewcommand\algorithmicensure{\textbf{Output:}}

\providecommand{\varnothing}{\emptyset}

\title{Multilayer Evaluation of Vertiport Departure Directions Under Urban Obstacles and Wind}

\author{Seunghyeop Jeong\footnote{Graduate Student, Graduate School of Automobile and Mobility; seunghyeopjeong@gmail.com.}}
\affil{Kookmin University, Seoul 02707, Republic of Korea}
\author{Max Z. Li\footnote{Assistant Professor, Department of Aerospace Engineering; maxzli@umich.edu.}}
\affil{University of Michigan, Ann Arbor, Michigan 48109}
\author{Kyowon Song\footnote{Associate Professor, Graduate School of Automobile and Mobility; kyowon@kookmin.ac.kr (Corresponding Author).}}
\affil{Kookmin University, Seoul 02707, Republic of Korea}

\begin{document}

\maketitle

\begin{abstract}
Vertiports must accommodate electric vertical takeoff and landing (eVTOL)
aircraft with different climb capabilities and crosswind tolerances. However,
conventional runway based departure criteria do not exploit the directional
freedom of vertical takeoff. This paper proposes a framework that evaluates
every departure azimuth and identifies usable and preferred departure directions
for each climb capability. Each azimuth is evaluated by its obstacle clearance
from surrounding urban buildings and its wind operability based on ten years of
wind observations. Three climb angle layers reflect differences in climb
capability among aircraft. Four weighting scenarios between obstacles and wind
reveal how operational priority affects the preferred departure direction. In an
application to Suseo station and the Jamsil heliport in Seoul, steeper climbs
widened the usable azimuths from 8.9 to 34.4\% and from 66.1 to 83.6\%,
respectively. By relaxing obstacle constraints, steeper climbs also shifted the
preferred direction toward the prevailing wind at lower wind weights. These
results show that vertiport departure paths should be planned as a set of
candidate directions matched to the obstacles and wind of the site and the climb
capability of the aircraft, rather than as a single runway style path.
\end{abstract}

\section*{Nomenclature}

{\renewcommand\arraystretch{1.0}
\noindent\begin{longtable*}{@{}l @{\quad=\quad} l@{}}
$A_{g}$ & directional availability of climb angle layer $g$ \\
$C_{i}$ & vertical margin of obstacle $i$ normalized by the local OCS height \\
${CM}_{i}$ & clearance margin of obstacle $i$ relative to the OCS, m \\
$D_{horizontal}$ & reference horizontal distance for the OCS slope, m \\
$d_{buffer}$ & distance from the centerline to the outer boundary of the lateral buffer zone, m \\
$d_{horizontal,i}$ & along track distance from the FATO edge reference point to obstacle $i$, m \\
$d_{obs}$ & lateral distance of an obstacle from the centerline, m \\
$d_{OCS}$ & half width of the core OCS corridor at the same along track distance, m \\
$f_{s}$ & occurrence frequency of compass sector $s$, h \\
$f_{\theta}$ & occurrence frequency of the compass sector containing azimuth $\theta$, h \\
$g$ & climb angle layer, deg \\
$H_{climb}$ & climb path height at $D_{horizontal}$, m \\
$H_{climb,i}$ & climb path height at the location of obstacle $i$, m \\
$H_{hov}$ & liftoff height at which the climb begins, m \\
$H_{OBS,i}$ & height of obstacle $i$, m \\
$H_{OCS}$ & OCS height at $D_{horizontal}$, m \\
$H_{OCS,i}$ & OCS height at the location of obstacle $i$, m \\
$i$ & obstacle index \\
$K$ & set of weighting scenarios \\
$k$ & weighting scenario \\
$L(\theta)$ & obstacles entering the lateral assessment of azimuth $\theta$ \\
$N_{g}^{usable}$ & number of usable azimuths in climb angle layer $g$ \\
$O(\theta)$ & obstacles entering the vertical assessment of azimuth $\theta$ \\
$P_{g,V_{cl}}^{(k)}$ & ranked candidate set of layer $g$ for crosswind limit $V_{cl}$ under scenario $k$ \\
$p$ & exponent of the lateral score \\
$q$ & exponent of the crosswind penalty \\
${ROC}_{i}$ & vertical room between the climb path and the OCS at obstacle $i$, m \\
$S_{ac}$ & long term mean crosswind operability \\
$S_{cw}$ & instantaneous crosswind score \\
$S_{HOR}$ & lateral obstacle clearance score \\
$S_{obs}$ & combined obstacle clearance score \\
$S_{pw}$ & prevailing wind score \\
$S_{total}^{(k)}$ & integrated score under weighting scenario $k$ \\
$S_{VER}$ & vertical obstacle clearance score \\
$S_{wind}$ & wind operability score \\
$s$ & compass sector index \\
$t$ & observation time \\
$V$ & observed wind speed, kt \\
$V_{cl}$ & crosswind limit, kt \\
$V_{cw}$ & crosswind component, kt \\
$W$ & ten year hourly wind record \\
$w_{obs}^{(k)}$ & obstacle weight of weighting scenario $k$ \\
$w_{wind}^{(k)}$ & wind weight of weighting scenario $k$ \\
$\beta$ & angular difference between the wind direction and the azimuth, deg \\
$\Theta$ & set of the 16 compass sectors \\
$\theta$ & candidate azimuth (departure heading), clockwise from true north, deg \\
$\theta_{climb}$ & climb angle of a layer, deg \\
$\theta_{OCS}$ & equivalent OCS slope angle, deg \\
$\tau$ & minimum integrated score for candidate selection \\
$\phi$ & observed wind direction, deg
\end{longtable*}}

\section{Introduction}
\label{sec:intro}

\lettrine{C}{hronic} congestion, environmental limits, and the structural
limits of existing ground infrastructure increasingly constrain urban transportation
systems, and this has created a need for new mobility modes \citep{holden2016fast}.
In this context, urban air mobility (UAM) based on electric vertical takeoff and
landing (eVTOL) aircraft has attracted attention as a new mode of transportation
that extends mobility into urban airspace at low altitude, and research has
accumulated on demand modeling, operations, and integration with existing
infrastructure \citep{garrow2021urban}. Institutional and operational
frameworks for actual operations and commercialization are also taking
shape \citep{faa2023conops}. In preparation for such operations, manufacturers
such as Joby Aviation, Archer Aviation, and EHang are racing to develop eVTOL
aircraft \citep{hu2025development}. An aircraft, however, cannot operate
without somewhere to take off and land, and \citet{sengupta2025uam} note that
UAM requires vertiport infrastructure, network planning, and air traffic
management to advance alongside vehicle technology. Among these, the vertiport
is the physical interface at which aircraft, airspace, and ground
infrastructure meet. Its departure and arrival paths are more than routes: they
govern the safety of the flight segments closest to the ground, and because
they set the turning and flight time needed to reach each destination, they
also shape how many aircraft the vertiport can handle. How these paths are
defined at the planning stage therefore shapes the operational capability of
the whole network.

\subsection{Fleet Heterogeneity and Vertiport Operations}
\label{subsec:heterogeneity}

One characteristic that distinguishes UAM operations from conventional air
transportation is that a single vertiport must accommodate aircraft with different
performance characteristics. The cruise performance and flight characteristics
of eVTOLs vary substantially with configuration and propulsion type, such as
multirotor, lift \& cruise, and tiltrotor; a review of UAM research classifies
these concepts into groups that differ fundamentally in cruise speed, range,
and hover efficiency \citep{straubinger2020overview}.

Previous work has shown how fleet heterogeneity affects operational
performance. \citet{kim2020receding} compared six fleet compositions and showed
that fleet heterogeneity affects both service quality and operational
efficiency and that the two conflict with each other. Fleet heterogeneity
extends beyond fleet size or seat count: differences in vehicle configuration
translate into operating parameters and constraints specific to each vehicle,
such as cruise speed, geometry, and energy
consumption \citep{husemann2024analysis}. In UAM, differences in vehicle
configuration are therefore a variable in operational planning, not only a
matter of aircraft design.

These differences also matter for terminal operations. Existing studies have
designed terminal airspace structures and arrival procedures while addressing
vertiport arrival management and
sequencing \citep{pradeep2020heuristic, bertram2020efficient, song2021development},
but they largely focus on how aircraft are arranged and operated within a given
path structure. A more fundamental planning question precedes this: when
aircraft with different climb capabilities and crosswind tolerances share a
single vertiport, which departure directions should be designated as usable,
and which of them are most favorable?
This question has received comparatively little attention.

\subsection{Departure Criteria and the Directional Freedom of Vertiports}
\label{subsec:transplant}

Vertiport approach and departure paths are currently assessed with obstacle
limitation surfaces (OLS), a concept carried over from aerodrome
standards \citep{icao2018annex14} that current vertiport guidance applies on the
premise of visual flight \citep{easa2022vertiports, faa2024eb105}. The Federal
Aviation Administration (FAA) vertiport design guidance specifies the geometry
of the approach/departure and transitional surfaces and treats a direction as
usable when no obstacle penetrates them \citep{faa2024eb105}. OLS defines the
airspace to be protected, while the path actually flown under visual
conditions is left to the pilot's visual judgment, so existing discussion of
vertiport departures has largely remained at the level of protection surfaces.

Instrument flight procedures (IFP), by contrast, divide the path into segments,
each with required obstacle clearance and navigation performance requirements,
independently of the pilot's visual judgment. Many conventional instrument
departure procedures, however, presuppose a unidirectional initial climb with
fixed geometry, a structure that originates in the physical constraint of a
runway \citep{faa2024terps, icao2018pansops}. Transplanting this structure
unchanged to a vertiport inherits proven safety principles but leaves vertical
takeoff and landing, the capability that most distinguishes eVTOLs from
conventional aircraft, unexploited at the procedure level.

\begin{figure}[htbp]
	\centering
		\includegraphics[width=0.44\linewidth]{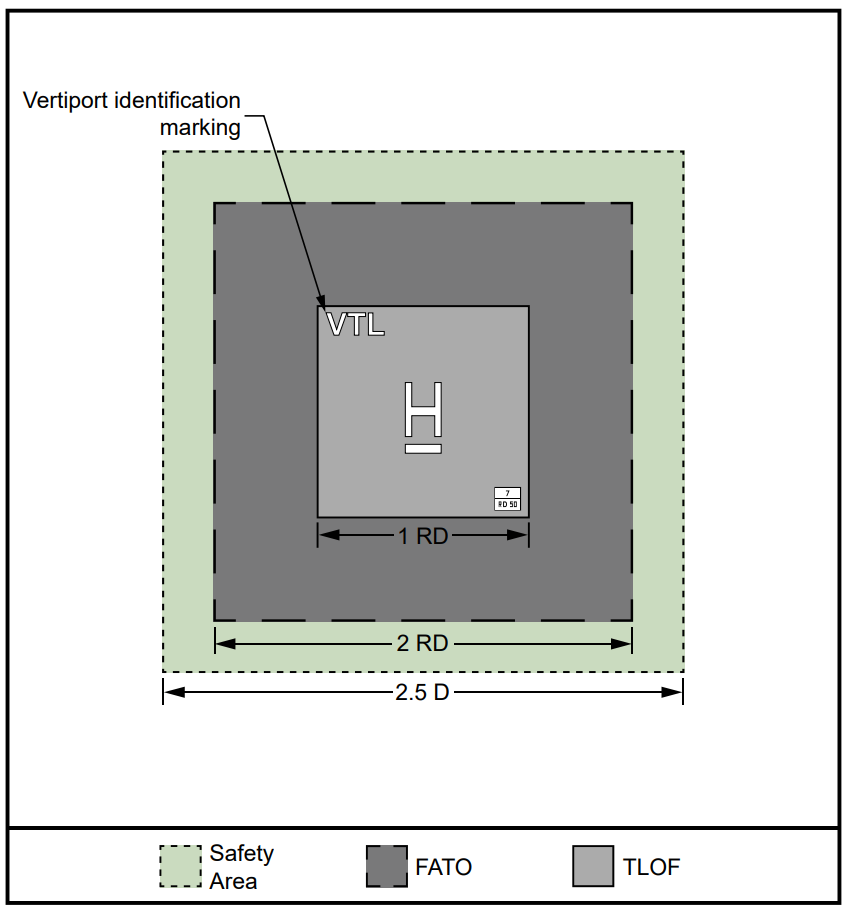}
	\caption{Vertiport takeoff and landing area, reproduced from FAA EB-105A \citep{faa2024eb105}: TLOF at least one rotor diameter (RD) wide, within the 2 RD FATO and safety area.}
	\label{fig:vertiport}
\end{figure}

The FAA vertiport design guidance lays out the takeoff and landing area as
three nested areas centered on a single point (Fig.~\ref{fig:vertiport}): a
touchdown and liftoff area (TLOF) at least one rotor diameter (RD) wide, on
which the aircraft touches down and lifts off; a final approach and takeoff
area (FATO) 2 RD wide around it, over which the approach is completed and from
which the takeoff begins; and a safety area 2.5 D wide, where D is the
controlling dimension of the design aircraft \citep{faa2024eb105}. Because the
aircraft lifts off vertically from the center of this layout, the facility
fixes no preferred direction, and departures can in principle use the full
directional domain. Unlike an airport, whose runway direction is set to suit
open land chosen at the planning stage, a vertiport is placed within an
existing urban fabric, so its usable directions must be found from the
surrounding obstacles and the preferable ones judged from the local wind.

The National Aeronautics and Space Administration (NASA) has introduced
omnidirectional use of the terminal airspace: the dynamic procedure design
concept of its eVTOL instrument flight procedure test plan envisions departure
and approach in any direction at a vertiport, with sectors withheld on the
basis of an obstacle evaluation \citep{zahn2024evtol}. This concept decides
whether a sector is usable but does not quantify the margin in each direction,
combine it with wind operability, or compare the result across climb profiles.
Existing site selection studies likewise apply static OLS that presuppose
visual flight, either in all directions at once \citep{brusberg2021landing} or
rotated around the site to find the directions free of
obstacles \citep{so2023integrated, guo2024vtol}, and judge each direction only
as usable or unusable without considering wind. Work that optimizes instrument
procedure design addresses conventional aviation and presupposes the
unidirectional structure of a runway \citep{andric2024novel}, and the influence
of wind has been reported for eVTOL operations and vertiport traffic
flow \citep{schweiger2023impact} without being combined direction by direction
with obstacle conditions.

\subsection{Transition to Predefined Criteria}
\label{subsec:gap}

This gap is also tied to how operations will change. As described above,
current UAM operational concepts mostly presuppose visual flight rules
(VFR) \citep{easa2022vertiports, faa2024eb105, kim2024design}, so that the
choice of departure direction and the judgment of wind conditions are delegated
to the pilot in real time. This premise, however, is unlikely to hold as
operations mature. The FAA UAM concept of operations states that initial UAM
operations will use the current regulatory framework of VFR and instrument
flight rules (IFR) as they are, that UAM corridors will be introduced as the
operational tempo increases, and that automated flight rules (AFR) are likely
to become available within the corridors as they become more defined. AFR is
presented as an evolution of the current VFR/IFR framework that reflects
advances in technology and procedures and supports higher capacity, and
vertiports are expected to serve VFR, IFR, and AFR operations
alike \citep{faa2023conops}. Other concept of operations documents likewise
envision a phased transition from piloted operations toward high density
autonomous and uncrewed operations \citep{embraerx2020, wisk2023conops}.

The detailed wind environment at an individual vertiport matters, but it belongs to the operations that follow the planning stage, where
it can be integrated with the wind of the moment. What the planning stage
requires is a framework that converts meteorological information into
directional usability. This study presents such a framework, and its output
describes, for each vertiport, which departure directions are usable, in a form
that higher level traffic management models can take as input.

\subsection{Research Objective and Contributions}
\label{subsec:contribution}

This leads to the central question of this study: when eVTOL aircraft with
different climb and crosswind capabilities share a single vertiport, how can
operational flexibility in the departure direction be secured? This study
addresses it by deriving the set of usable directions for each climb angle layer
instead of proposing one optimal departure direction. The obstacle clearance
assessment borrows principles established in instrument procedure design, while
the evaluation domain and structure are built on the omnidirectional egress and
multilayer structure of a vertiport.

The main contributions are as follows. First, a framework is proposed that
evaluates vertiport departure directions over the entire directional domain in
a dense urban obstacle environment and quantifies the obstacle clearance margin
in each direction, in addition to a usable or unusable judgment. Second, the
framework integrates this margin with operability derived from long term wind
records, so that the wind factor is quantified in advance within the direction
evaluation, as the planning stage requires. Third, the set of usable directions
and their margins are derived for three climb angle layers and three crosswind
tolerance scenarios at two Seoul sites with contrasting surroundings, and the
sensitivity of the candidate rankings to the priority between obstacles and wind
is analyzed. The obstacle clearance geometry and the wind thresholds can be
replaced as vertiport specific standards emerge, and further criteria such as
energy consumption or passenger comfort can be added within the same structure.

\section{Structure of the Vertiport Departure Path}
\label{sec:structure}

\subsection{Adoption of Obstacle Clearance Principles}
\label{subsec:principles}

Since no obstacle clearance standard specific to vertiports has yet been
established, this study borrows the principles used in instrument procedure
design. The two central concepts are the obstacle clearance surface (OCS), a
sloped protection surface above which the aircraft must remain, and the
required obstacle clearance (ROC), the minimum vertical margin that must be
secured from obstacles \citep{faa2024terps, icao2018pansops}. These principles
apply with particular force to the initial departure segment, where the
aircraft is still at low altitude and close to the surrounding obstacles, and
for the same reason the initial departure segment is the focus of this
evaluation.

IFP criteria are adopted here for the following reason. The rule framework that
will govern highly automated eVTOL operations, the AFR described in
Sec.~\ref{subsec:gap}, exists only as an operational concept and has not been
defined as an applicable design standard. VFR criteria leave obstacle clearance and
wind assessment to the pilot in real time and do not provide the same segment
specific obstacle clearance framework as IFP criteria. IFP criteria express
obstacle clearance as explicit, quantitative surfaces that do not depend on
visual judgment, and this property is what allows them to be combined with
quantified wind criteria within a single evaluation structure. This study
therefore borrows IFP criteria as an analytical basis for planning stage
evaluation.

\begin{figure}[htbp]
	\centering
		\includegraphics[width=0.6\linewidth]{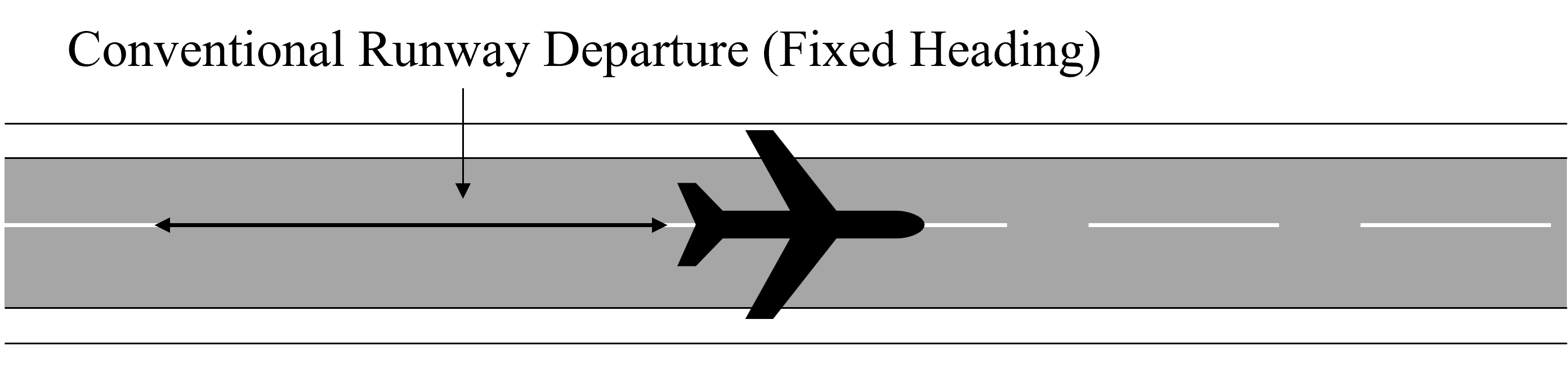}
	\caption{Single runway operation of conventional aircraft.}
	\label{fig:runway}
\end{figure}

The conditions under which these principles apply, however, differ
considerably from those at conventional airports. As Fig.~\ref{fig:runway}
shows, runway operations presuppose a unidirectional departure with fixed
geometry. A vertiport has no physical directional constraint, and dense urban
obstacles, rooftop structures, and a highly variable wind environment determine
which of its directions are usable. Even when the same clearance principles are
applied, the result should therefore take the form of a distribution of
directional usability rather than a pass/fail judgment for a single direction.

Several recent studies and regulatory documents likewise present structures
that use the full directional domain in urban settings: the terminal control
model based on concentric rings \citep{song2021development}, the concentric
arrival structure \citep{bertram2020efficient}, and the obstacle free volumes
(OFV) introduced by the European Union Aviation Safety Agency (EASA), under
which omnidirectional trajectories are possible \citep{easa2022vertiports}.

\subsection{Multilayer Structure for Heterogeneous Fleets}
\label{subsec:layers}

\begin{figure}[htbp]
	\centering
		\includegraphics[width=0.83\linewidth]{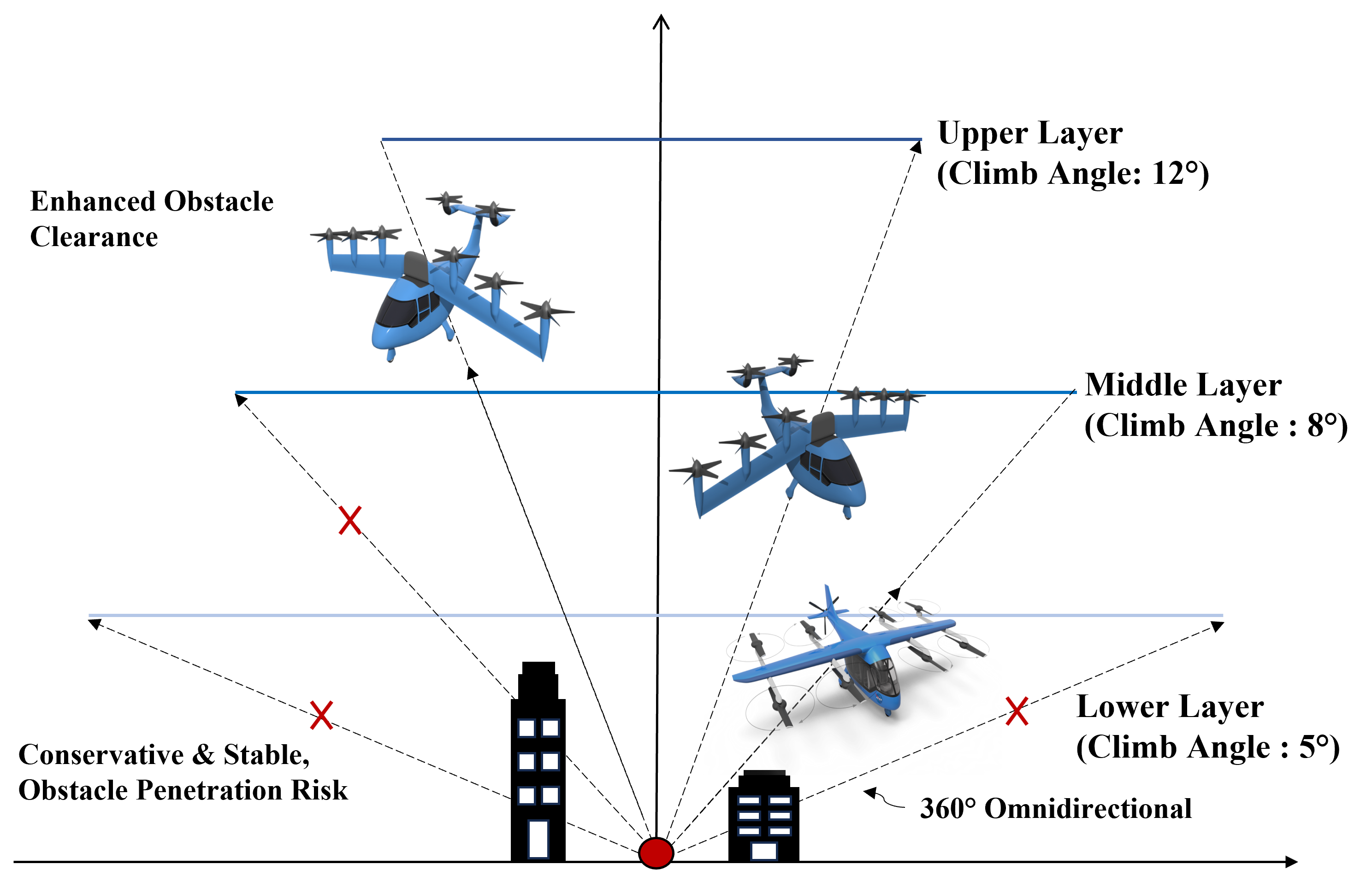}
	\caption{Multilayer departure path structure composed of $5^{\circ}$,
$8^{\circ}$, and $12^{\circ}$ climb angles.}
	\label{fig:layers}
\end{figure}

To accommodate aircraft of different performance at a single vertiport, this
study assumes a multilayer departure path structure composed of three climb
angles, $5^{\circ}$, $8^{\circ}$, and $12^{\circ}$ (Fig.~\ref{fig:layers}).
Aircraft requiring a shallow climb can use the lower layer and aircraft capable
of a steep climb the higher layer, which widens the range of fleet compositions
a vertiport can serve compared with forcing one path geometry on all types.
Ideally the departure path radius should vary with the climb angle to reflect
the different climb profiles, but this study simplifies the geometry by
applying the same radius, the $3038\,\mathrm{ft}$ ($926\,\mathrm{m}$) initial segment of
Sec.~\ref{subsec:lateral}, to all layers.

The three angles are taken from the evaluation of candidate eVTOL instrument
flight procedures that the NASA Advanced Air Mobility National Campaign
conducted with Joby Aviation. In that evaluation, vertical profiles of
$5^{\circ}$, $8^{\circ}$, and $12^{\circ}$ were flown for the departure, final
approach, and missed approach segments in Joby's engineering flight simulator,
with safety, efficiency, and passenger comfort among the assessment
criteria \citep{zahn2023uam}, and the same three angles were carried into the
vehicle agnostic test plan that followed, which also lists them in its example
departure sequence \citep{zahn2024evtol}. At the time of writing, no piloted
passenger eVTOL has completed type certification with the FAA or EASA, so
certified climb performance is not available for any of these aircraft. In its
absence, this study adopts the three angles that were flown as representative
path angles in these simulator evaluations of a specific eVTOL design.

Once Joby and other types complete certification and their performance becomes
available, the number and angles of the climb layers can be revised against
factors such as battery energy consumption and passenger comfort, and the
framework accepts such values without any change to its structure. Because the
purpose here is to show how the set of usable directions changes with the
allowable climb profile, three layers spanning shallow to steep climbs suffice,
as the results for both case study sites in Sec.~\ref{subsec:obsresult}
confirm.

The layers are evaluated separately; departing on the same azimuth in several
layers at once would require vertical separation standards and traffic
management rules of its own. What this study presents is an evaluation of which
azimuths are appropriate for departure in each climb angle layer. The output is thus the usability and priority of
climb angle and azimuth combinations, which can serve as input to the later
design of operating rules and the detailing of procedures.

\subsection{Evaluation Structure}
\label{subsec:evalstructure}

On these premises, the evaluation structure of this study differs from the
conventional approach in two respects. Every azimuth over $360^{\circ}$ is
evaluated instead of a single fixed direction, and the size of the margin is
evaluated alongside the pass/fail judgment. Two azimuths that equally satisfy
the obstacle clearance requirement should not be treated as equivalent if one
lies much closer to the protection surface, and this distinction is what allows
the many usable directions to be ranked for further examination.

\section{Data Collection}
\label{sec:data}

To construct the reference dataset for the directional evaluation, this study
integrated obstacle and meteorological data. Building data including the
number of stories were taken from the digital topographic data of the National
Geographic Information Institute (NGII) \citep{ngii2025}. Verified height values
for individual buildings are not publicly released, so height was estimated at
$4\,\mathrm{m}$ ($13.1\,\mathrm{ft}$) per story. Terrain elevation was not
included, so the obstacle height is the building height alone. The building layer was
preprocessed in QGIS under a unified coordinate system and used
for the obstacle penetration assessment.

Meteorological data were obtained from the hourly records of the automated
synoptic observing system (ASOS) station in Seoul (station 108) operated by the
Korea Meteorological Administration, covering the ten years from July 2015 to
June 2025 \citep{kma2025}, the record length of ten consecutive years that
FAA airport design guidance prefers for wind
analysis \citep{faa2024airportdesign}. The station lies about $15\,\mathrm{km}$ from
Suseo station and about $11\,\mathrm{km}$ from the Jamsil heliport, and the
same record is used as a proxy for the local wind environment at both sites.
Wind direction in this record is reported in 16 compass points coded in whole
degrees, and direction code 0 denotes calm conditions. Each reported direction
is assigned to the $22.5^{\circ}$ sector of its compass point
(Sec.~\ref{sec:wind}), and wind speed, reported in $\mathrm{m\,s^{-1}}$, is
converted to knots ($1\,\mathrm{m\,s^{-1}} \approx 1.944\,\mathrm{kt}$) before it
is compared with the crosswind limits. After the hours with a missing speed or
direction are removed, the record contains 87,309 hourly observations, of which
2,605 (about 3\%) are calm. The detailed scoring of the prevailing wind and
crosswind is described in Sec.~\ref{sec:wind}.

This integrated dataset feeds an analysis model implemented in the Python
console of QGIS, which evaluates obstacle clearance and directional usability
automatically. The results are then passed to an external scoring routine that
produces the evaluation for each azimuth.

\begin{figure}[htbp]
	\centering
		\includegraphics[width=\linewidth]{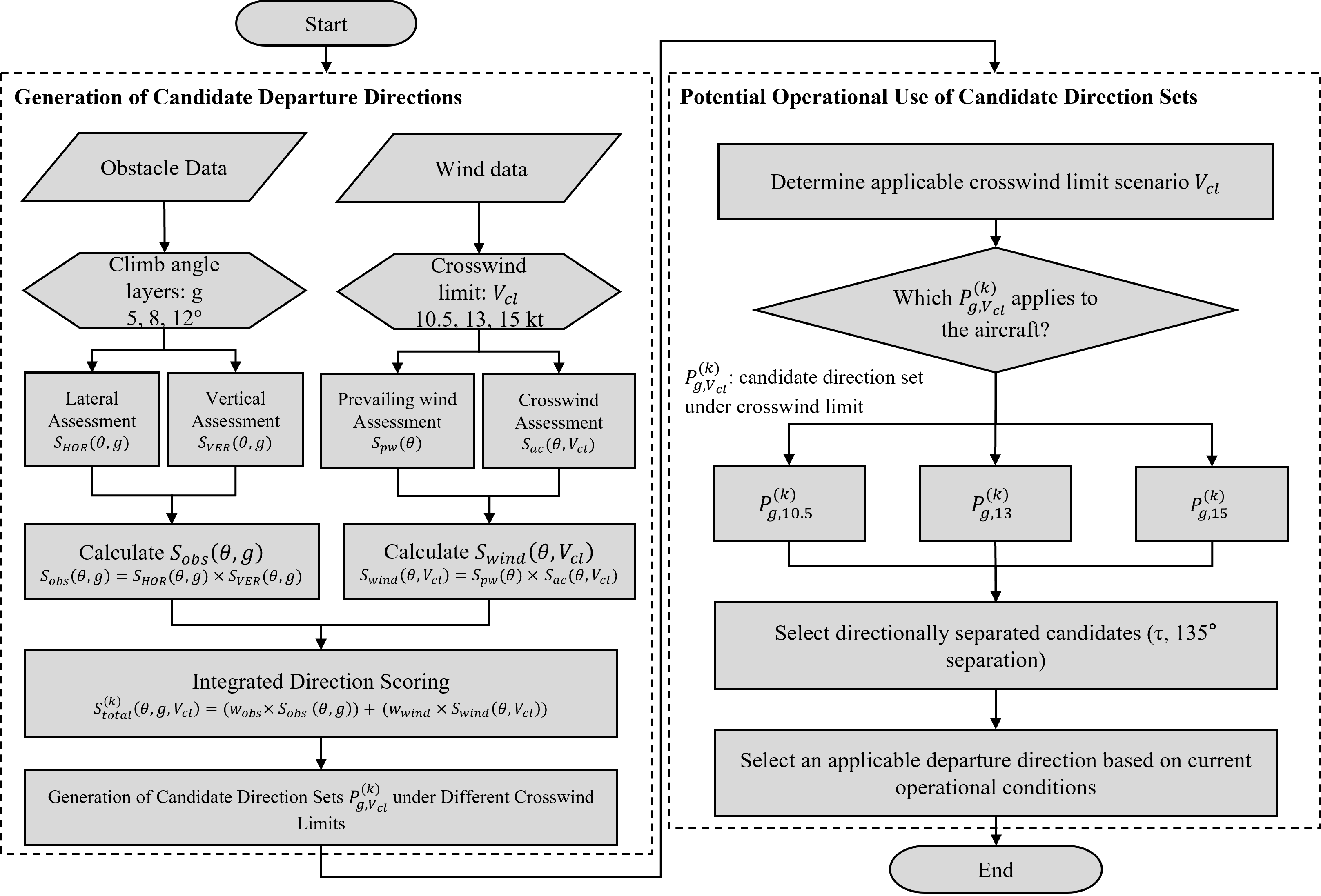}
	\caption{Overall workflow of the proposed evaluation framework.}
	\label{fig:workflow}
\end{figure}

\section{Direction Evaluation Framework}
\label{sec:framework}

The proposed framework evaluates and ranks vertiport departure directions from
the margin available in each direction, drawing on the obstacle clearance
principles described in Sec.~\ref{subsec:principles} and on wind operability
conditions. The resulting scores express the relative suitability of azimuths,
and the overall workflow is shown in Fig.~\ref{fig:workflow}. The symbols used
in the framework are listed in the Nomenclature.

The evaluation follows a staged structure that progressively narrows the
candidates from geometric usability to directional operability, and is
performed independently for each climb angle layer. The lateral obstacle
assessment acts first as a strict usability filter, excluding azimuths that
violate the protected corridor defined by required navigation performance
(RNP). A vertical assessment of the obstacle clearance margin is then performed
for the remaining azimuths. The vertical assessment is described first
(Sec.~\ref{subsec:vertical}) because the lateral assessment uses the protection
surface height defined there. The two results are combined into a single
obstacle clearance score, which forms the core of the framework and is applied
across the climb angle layers (Sec.~\ref{sec:application}).

\subsection{Vertical Obstacle Assessment}
\label{subsec:vertical}

\begin{figure}[H]
	\centering
		\includegraphics[width=\linewidth]{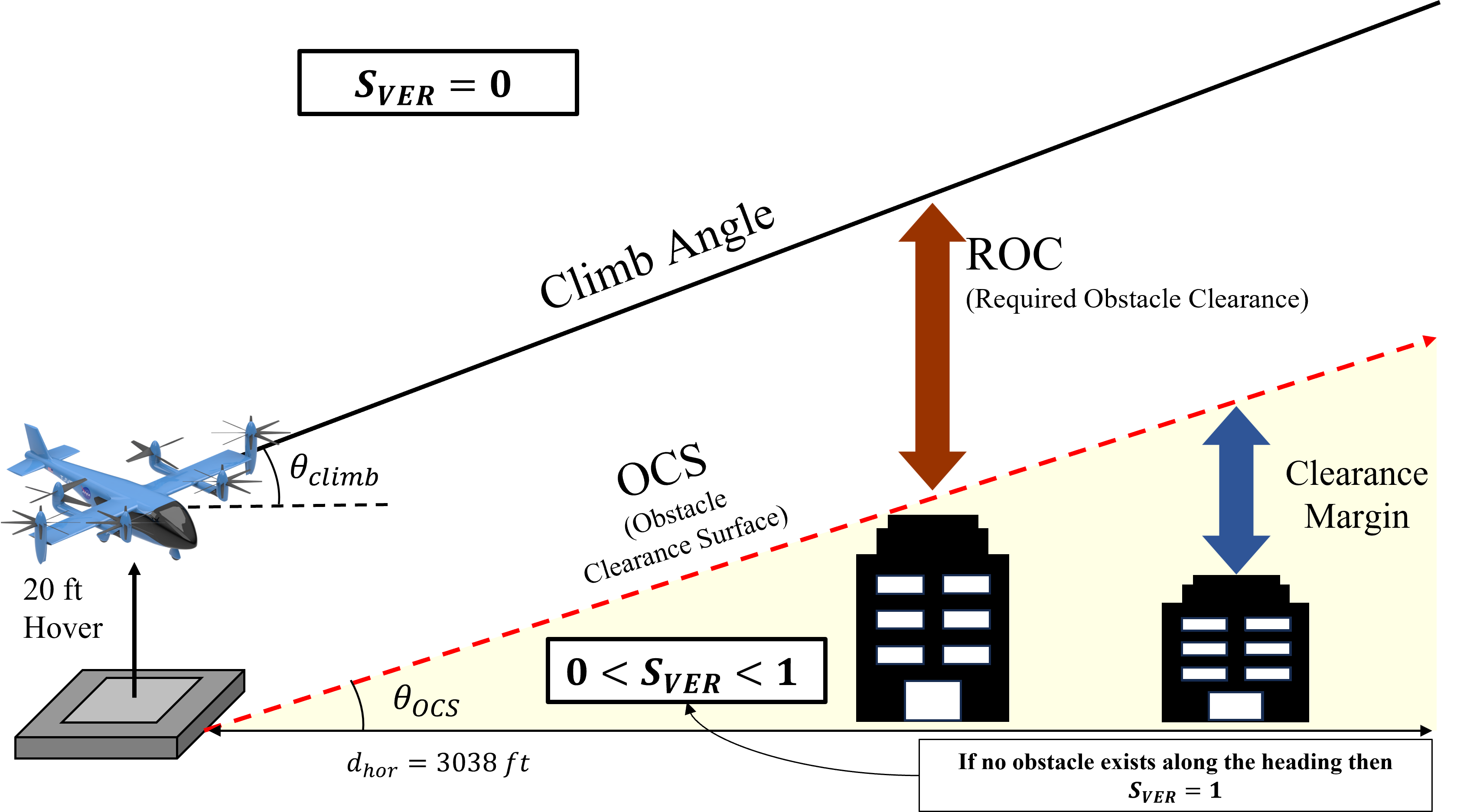}
	\caption{Geometry of the vertical obstacle assessment.}
	\label{fig:vertgeom}
\end{figure}

Conventional obstacle assessment focuses mainly on judging whether an obstacle
penetrates the OCS. This study also evaluates how much margin remains even when
no penetration occurs, because an obstacle immediately below the protection
surface cannot be treated the same as one far below it. The climb path height
is the primary reference for this margin based evaluation, and climb angles of
$5^{\circ}$, $8^{\circ}$, and $12^{\circ}$ are considered
(Sec.~\ref{subsec:layers}). In every layer the aircraft is assumed to lift off
vertically and begin its climb at $20\,\mathrm{ft}$ ($6.1\,\mathrm{m}$). The adopted geometry is
shown in Fig.~\ref{fig:vertgeom}.

Instead of the geometric center of the FATO, a reference point on the FATO edge
is adopted to reflect the actual initiation point of the departure path.
Because the FATO dimensions depend on the design aircraft \citep{faa2024eb105},
both case studies assume a rotor diameter of $15.2\,\mathrm{m}$
($50\,\mathrm{ft}$) and hence a square FATO $30.4\,\mathrm{m}$ (2 RD) wide.
The reference point then lies $15.2\,\mathrm{m}$ ($50\,\mathrm{ft}$) from the center; the FATO
edge is approximated by a circle of this radius, so the reference point lies at
the same distance for every azimuth. The angular sector within which each
obstacle can affect the departure corridor is defined from this reference
point. The azimuth toward the leftmost edge of an obstacle is defined as its
start angle and that toward its rightmost edge as its end angle, and an
obstacle lies on the centerline of an azimuth when the sector between these two
angles contains it. The vertical assessment of an azimuth covers the obstacles
on its centerline and every other obstacle within its core corridor
(Sec.~\ref{subsec:lateral}), because the aircraft may be anywhere within that
corridor. Azimuths are measured clockwise from true north at this reference
point and denote the direction in which an aircraft departs from the FATO,
which is also its outbound course. ``Heading'' in the figures that follow refers to
this azimuth.

The OCS slope is defined with a reference horizontal distance $D_{horizontal}$
of $1\,\mathrm{NM}$ ($6076.12\,\mathrm{ft}$, $1852\,\mathrm{m}$). Because the climb path slope
differs from the OCS slope, an equivalent OCS slope angle is derived for each
configuration. Here $H_{OCS}$ denotes the OCS height at the reference
horizontal distance, and the climb path height at that location is computed
first in order to obtain the OCS slope angle. The climb path height is computed
from the assumed climb angle and a liftoff height of $20\,\mathrm{ft}$, which
lies within the range of initial hover and ground effect heights
($10$--$20\,\mathrm{ft}$, $3$--$6\,\mathrm{m}$) in the departure profiles of \citet{zahn2024evtol}.
\begin{equation}
H_{climb} = H_{hov} + D_{horizontal} \times \tan\left( \theta_{climb} \right)
\label{eq:hclimb}
\end{equation}
where $H_{climb}$ is the climb path height at $D_{horizontal}$, $H_{hov}$ the
liftoff height, and $\theta_{climb}$ the climb angle of the layer. The OCS height at the reference horizontal distance is defined as a fixed
fraction of the corresponding climb path height.
\begin{equation}
H_{OCS} = 0.76 \times H_{climb}
\label{eq:hocs}
\end{equation}
The coefficient 0.76 in Eq.~\eqref{eq:hocs} follows the departure obstacle
evaluation criterion of the NASA vehicle agnostic eVTOL flight procedure test
plan, which applies a 76\% OCS to terrain and obstacles; in the same plan's
glide slope guidance, the remaining 24\% is the clearance between the flight
path and the OCS \citep{zahn2024evtol}.

Using the derived OCS height, the equivalent OCS slope angle is computed as
follows.
\begin{equation}
\theta_{OCS} = \arctan\left( \frac{H_{OCS}}{D_{horizontal}} \right)
\label{eq:thetaocs}
\end{equation}
Applying this equation yields equivalent OCS angles of $3.95^{\circ}$,
$6.24^{\circ}$, and $9.32^{\circ}$ for climb angles of $5^{\circ}$,
$8^{\circ}$, and $12^{\circ}$, respectively. To reflect the $20\,\mathrm{ft}$ liftoff
height from which an eVTOL begins its climb, this study includes that height in
the OCS height at $1\,\mathrm{NM}$ when deriving the equivalent slope, and
computes the OCS height at each obstacle location from the linear surface that
rises at this slope from the reference point (Eq.~\eqref{eq:hocsi}).

The resulting OCS angles are used in an obstacle evaluation routine written in
Python. Although the OCS slope is derived at the $1\,\mathrm{NM}$ reference
distance, the obstacle evaluation itself is performed only within the
$3038\,\mathrm{ft}$ initial departure segment described in
Sec.~\ref{subsec:lateral}; obstacles beyond this distance are not considered.
For each obstacle, the horizontal distance $d_{horizontal,i}$ is measured from
the same FATO edge reference point along the climb path azimuth to the
projected position of the obstacle centroid, and the OCS height at the obstacle
location is computed as follows.
\begin{equation}
H_{OCS,i} = d_{horizontal,i} \times \tan\left( \theta_{OCS} \right)
\label{eq:hocsi}
\end{equation}
The vertical room between the climb path and the protection surface at the
obstacle location is evaluated as follows. It is reported as the margin
available at that point and is not used as a decision variable.
\begin{equation}
{ROC}_{i} = H_{climb,i} - H_{OCS,i}
\label{eq:roci}
\end{equation}
where $H_{climb,i}$ is the climb path height at the location of obstacle $i$.
The decision is taken against the OCS itself: if the height of any obstacle
reaches or exceeds the OCS height at its location, the departure direction is
classified as unusable and excluded from further analysis. For obstacles that
remain below the OCS, a clearance margin ${CM}_{i}$ is defined to quantify the
vertical proximity to the surface at that location.
\begin{equation}
{CM}_{i} = H_{OBS,i} - H_{OCS,i}
\label{eq:cm}
\end{equation}
where $H_{OBS,i}$ is the height of obstacle $i$. A zero or positive value means that the obstacle reaches or penetrates the OCS,
and a negative value indicates a secured margin below the OCS. Unlike the pass/fail
test above, this margin is used directly in the vertical scoring to distinguish
azimuths according to obstacle proximity.

\textit{Vertical obstacle clearance score.} To quantify the vertical condition
of each candidate azimuth, the obstacle height $H_{OBS}$ is compared with the
OCS and climb path geometry. The OCS height is computed locally for each
obstacle location, and the vertical score of an azimuth is governed by the most
critical obstacle on it.

If any obstacle entering the vertical assessment reaches or exceeds the OCS
height at its location, the azimuth is classified as unusable and the vertical
score $S_{VER}$ is set to 0.
\begin{equation}
H_{OBS,i} \geq H_{OCS,i} \Rightarrow S_{VER} = 0
\label{eq:verzero}
\end{equation}
Even when no penetration occurs, treating an obstacle very close to the OCS in
the same way as one with ample margin overestimates the available margin. The
vertical margin of each obstacle is therefore normalized by the OCS height at
its own location.
\begin{equation}
C_{i} = \frac{H_{OCS,i} - H_{OBS,i}}{H_{OCS,i}}, \quad H_{OBS,i} < H_{OCS,i}
\label{eq:cmin}
\end{equation}
Here $C_{i}$ is the share of the protection surface height at that distance that
remains as margin, so that obstacles close to the FATO, where the surface is
low, and obstacles farther out are compared on the same scale. Since multiple
obstacles may lie on one azimuth, the obstacle with the smallest normalized
margin governs.
\begin{equation}
S_{VER} =
\begin{cases}
0, & H_{OBS,i} \geq H_{OCS,i} \text{ for any } i \\[2pt]
1, & \text{no obstacle} \\[2pt]
\min_{i} C_{i}, & \text{otherwise}
\end{cases}
\label{eq:sver}
\end{equation}
The case $0 < S_{VER} < 1$ corresponds to azimuths that are geometrically
usable but contain obstacles with limited vertical margin, with the score
decreasing as the available margin shrinks. $S_{VER} = 1$ indicates that no
obstacle enters the vertical assessment of the azimuth.

\subsection{Lateral Obstacle Assessment}
\label{subsec:lateral}

\begin{figure}[htbp]
	\centering
		\subfloat[OLS defined in FAA EB-105A]{\includegraphics[width=0.44\linewidth]{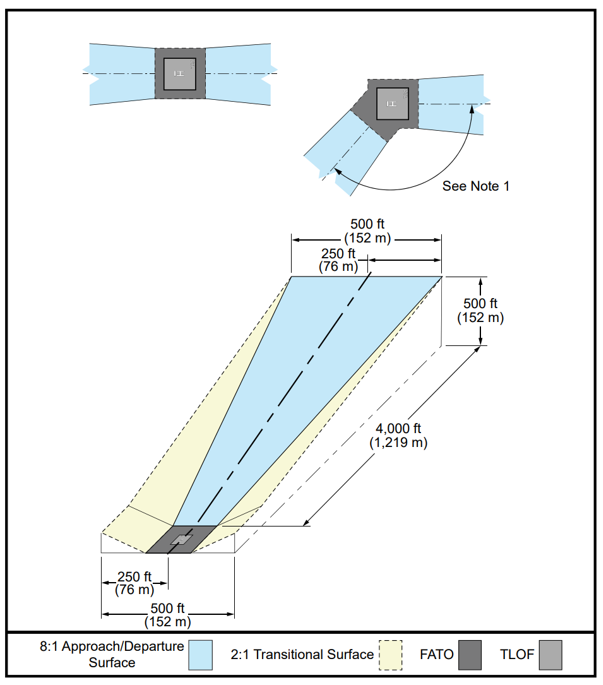}}\\[4pt]
		\subfloat[OCS with lateral buffer zones, based on the candidate RNP 0.1 surfaces of NASA AAM-NC-121-001]{\includegraphics[width=0.8\linewidth]{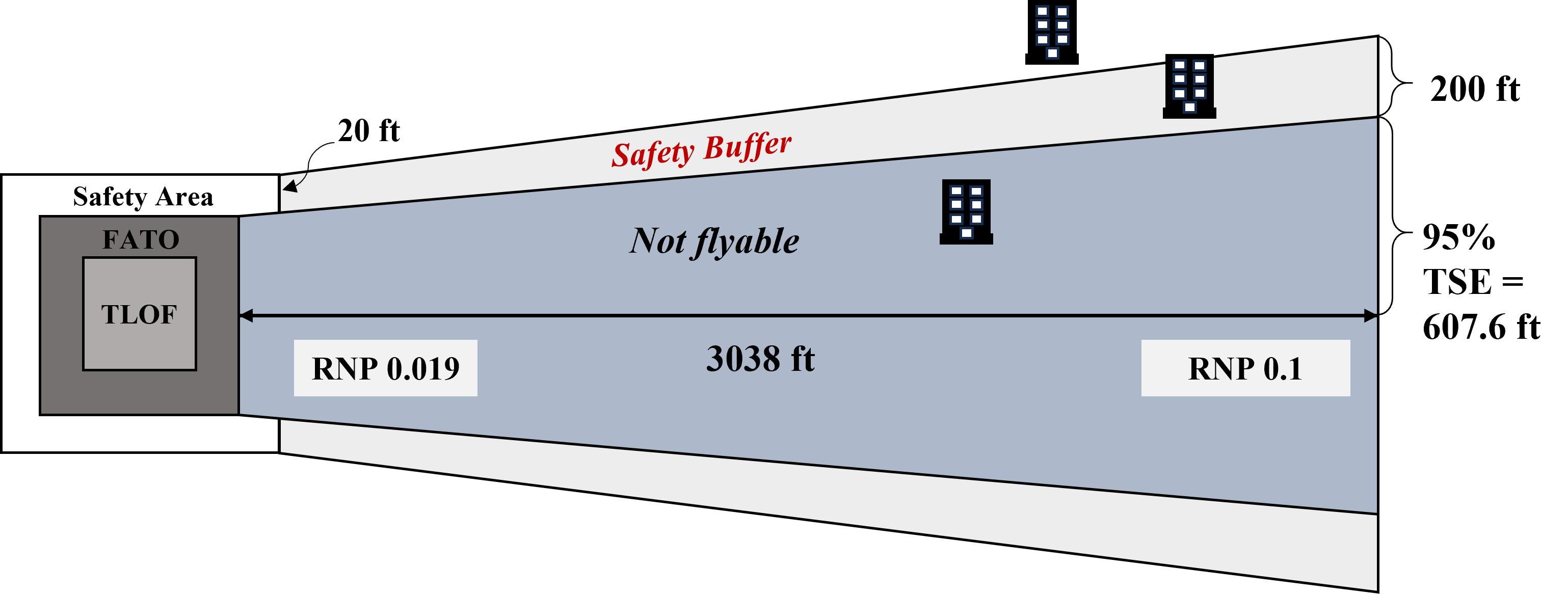}}
	\caption{VFR based OLS and RNP based protection surface.}
	\label{fig:ols}
\end{figure}

Previous studies applied OLS criteria derived from VFR, consisting of approach
and transitional surfaces. Operations oriented toward autonomy will rely on
navigation performance rather than visual confirmation, which calls for a
different evaluation structure. This study therefore evaluates the path against
RNP, in which the required navigation accuracy is specified as a performance
requirement \citep{icao2023pbn}, and the transitional surface is not used here.
Instead, lateral buffer zones are applied on both sides of the OCS. The lateral
corridor geometry adopted here is taken from the candidate RNP 0.1 surfaces
that \citet{zahn2024evtol} propose. Because this buffer is not a
mandatory regulatory requirement, its effect is handled through normalization
within the scoring process as described below. Figures~\ref{fig:ols}a and
\ref{fig:ols}b show, respectively, the conventional OLS with the transitional
surface \citep{faa2024eb105} and the OCS with lateral buffer zones used under
RNP.

The OCS expands with distance from the FATO, and at the terminal distance of
$3038\,\mathrm{ft}$ the corridor reaches the width that the test plan applies
for RNP 0.1: a 95\% total system error of $\pm 0.1\,\mathrm{NM}$
($\approx \pm 607.6\,\mathrm{ft}$ or $\pm 185.2\,\mathrm{m}$), that is, a full width of
$1215\,\mathrm{ft}$ ($370\,\mathrm{m}$) \citep{zahn2024evtol}. At the reference point the core
corridor is as wide as the FATO, $30.4\,\mathrm{m}$ (Sec.~\ref{subsec:vertical}),
in line with the FAA vertiport guidance, in which the departure surface begins
at the FATO edge with the width of the FATO \citep{faa2024eb105}; its half
width $d_{OCS}$ therefore increases linearly from $15.2\,\mathrm{m}$
($50\,\mathrm{ft}$) there to $607.6\,\mathrm{ft}$ at $3038\,\mathrm{ft}$. The
lateral buffer keeps a margin outside the core corridor; it adds
$20\,\mathrm{ft}$ ($6.1\,\mathrm{m}$) on each side at the reference point and widens linearly to
$200\,\mathrm{ft}$ ($61\,\mathrm{m}$) at $3038\,\mathrm{ft}$.

\textit{Lateral obstacle clearance score.} The lateral assessment considers the
obstacles that lie within the buffer zone on either side of the centerline and
whose height reaches the protection surface at their along track distance,
that is, $H_{OBS,i} \geq H_{OCS,i}$ with $H_{OCS,i}$ from Eq.~\eqref{eq:hocsi}.
It does not measure a vertical margin; the height only decides which obstacles
are relevant to the lateral assessment. For each such obstacle, the lateral
distance from the centerline ($d_{obs}$) is compared with the half width of the
core OCS corridor ($d_{OCS}$) and the outer edge of the lateral buffer zone
($d_{buffer}$), both measured from the centerline at the same along track
distance. The normalized lateral score is defined as follows.
\begin{equation}
S_{HOR} =
\begin{cases}
0, & d_{obs} < d_{OCS} \\[4pt]
\left( \dfrac{d_{obs} - d_{OCS}}{d_{buffer} - d_{OCS}} \right)^{p},
  & d_{OCS} \leq d_{obs} < d_{buffer} \\[8pt]
1, & d_{obs} \geq d_{buffer}
\end{cases}
\label{eq:shor}
\end{equation}
$S_{HOR} = 0$ means that an obstacle reaching the protection surface lies
within the core OCS corridor, and the azimuth is not flyable. Intermediate
values correspond to obstacles outside the core corridor but within the lateral
buffer zone, normalized by their proximity to the OCS boundary. $S_{HOR} = 1$
means that the obstacle lies entirely outside the buffer zone.

The intermediate region is modeled by a parameterized power law normalization,
in which the exponent $p$ controls how quickly the score decays near the OCS
boundary. A smaller $p$ softens the penalty for azimuths close to the boundary,
and a larger $p$ imposes a more conservative penalty. This study adopts
$p = 2$, which penalizes azimuths with limited margin while still
distinguishing among usable ones, and $p$ affects only the ordering of
candidates, not usability itself. Usability is decided by the strict condition
of Eq.~\eqref{eq:shor} and is independent of the choice of $p$.

\subsection{Combined Obstacle Clearance Score}
\label{subsec:combined}

The vertical and lateral results are combined into a single obstacle clearance
score $S_{obs}(\theta,g)$ of azimuth $\theta$ in climb angle layer $g$ using a
multiplicative formulation, as summarized in
Algorithm~\ref{alg:obs}.
\begin{equation}
S_{obs}(\theta,g) = S_{HOR}(\theta,g) \times S_{VER}(\theta,g)
\label{eq:sobs}
\end{equation}
\begin{algorithm}[htbp]
\caption{Obstacle clearance scoring by climb angle layer and azimuth.}
\label{alg:obs}
\small
\begin{algorithmic}[1]
\Require obstacle set $OBS$; climb angle layers $g \in \{5,8,12\}^{\circ}$;
  candidate azimuths $\theta \in \{0,1,\ldots,359\}^{\circ}$
\Ensure obstacle clearance score $S_{obs}(\theta,g)$
\ForAll{$g$}
  \For{$\theta = 0^{\circ}$ \textbf{to} $359^{\circ}$}
    \State $O(\theta) \gets$ obstacles in the $3038\,\mathrm{ft}$ initial segment
      on the centerline along $\theta$ or within $d_{OCS}$ of it
    \State $L(\theta) \gets$ obstacles in the initial segment within $d_{buffer}$
      of the centerline along $\theta$ with $H_{OBS,i} \geq H_{OCS,i}$
      (Eq.~\eqref{eq:hocsi})
    \If{$O(\theta) = \varnothing$ \textbf{and} $L(\theta) = \varnothing$}
      \State $S_{obs}(\theta,g) \gets 1$; \textbf{continue}
    \EndIf
    \Statex \hspace{\algorithmicindent}\emph{Step 1: lateral assessment}
    \State $S_{HOR} \gets 1$
    \ForAll{$i \in L(\theta)$}
      \State compute $d_{obs,i}$ and the local $d_{OCS}$ and $d_{buffer}$ at $d_{horizontal,i}$
      \If{$d_{obs,i} < d_{OCS}$} \State $S_{HOR} \gets 0$; \textbf{break}
      \ElsIf{$d_{obs,i} < d_{buffer}$}
        \State $S_{HOR} \gets \min\left( S_{HOR}, \text{Eq.~\eqref{eq:shor}} \right)$
      \EndIf
    \EndFor
    \If{$S_{HOR} = 0$} \State $S_{obs}(\theta,g) \gets 0$; \textbf{continue}
    \EndIf
    \Statex \hspace{\algorithmicindent}\emph{Step 2: vertical assessment}
    \ForAll{$i \in O(\theta)$}
      \State compute $H_{OCS,i}$ and ${CM}_{i}$ by Eqs.~\eqref{eq:hocsi} and \eqref{eq:cm}
    \EndFor
    \If{$O(\theta) = \varnothing$}
      \State $S_{VER} \gets 1$
    \ElsIf{$\exists\, i \in O(\theta): {CM}_{i} \geq 0$}
      \State $S_{VER} \gets 0$
      \Comment{obstacle at or above the OCS}
    \Else
      \State $S_{VER} \gets \min_{i \in O(\theta)} C_{i}$
        by Eqs.~\eqref{eq:cmin} and \eqref{eq:sver}
    \EndIf
    \Statex \hspace{\algorithmicindent}\emph{Step 3: combined score}
    \State $S_{obs}(\theta,g) \gets S_{HOR} \times S_{VER}$
  \EndFor
\EndFor
\end{algorithmic}
\end{algorithm}

Here $S_{HOR}(\theta,g)$ functions as the primary filter that checks whether
the core corridor is clear. Because the lateral assessment considers only
obstacles that reach the protection surface, the lateral margin is evaluated on
the same sloped surface that governs vertical usability rather than on a purely
horizontal plane; obstacles in the core corridor that remain below that surface
enter the score through the vertical assessment, and those in the buffer zone
below it do not affect the score. When an obstacle penetrates the core OCS
corridor and $S_{HOR}(\theta,g) = 0$, the combined score is 0 regardless of the
vertical margin and the azimuth is excluded.

For azimuths that pass the lateral test, the vertical score is applied to yield
the final margin, which classifies azimuths as unusable ($S_{obs} = 0$), margin
limited ($0 < S_{obs} < 1$), or unobstructed ($S_{obs} = 1$). The
multiplicative form was adopted because the lateral and vertical margins are
independent requirements and a large vertical margin cannot compensate for an
obstacle inside the core corridor.

The vertical margin is normalized against the OCS geometry of its own layer, so
$S_{obs}(\theta,g)$ is a relative measure within a layer. A higher score in the
$12^{\circ}$ layer than in the $8^{\circ}$ layer therefore indicates a wider
margin relative to that layer's own protection surface. Scores are compared
between azimuths within a layer, and between layers only in terms of which
azimuths are usable.

\subsection{Wind Operability Score}
\label{sec:wind}

Conventional runway planning relies on long term wind records. A runway orientation
is chosen so that the crosswind component stays below the allowable limit for
at least 95\% of the year, with allowable crosswind components of $10.5$, $13$,
$16$, and $20\,\mathrm{kt}$ depending on the runway design
code \citep{faa2024airportdesign}. The 95\% criterion exists because a runway
fixes the departure direction for its whole service life: with only one or at
most two directions available, a single orientation has to cover nearly every
wind condition on its own.

A vertiport is not bound in this way. Because departures can be made along any
usable direction, an azimuth that is unfavorable under the crosswind of a given
moment can be replaced at that moment by another usable azimuth, so that wind
coverage is provided by the set of usable directions as a whole. What the
planning stage needs from the wind record is thus a graded measure of how
favorable each azimuth is over the long term, and every azimuth over
$360^{\circ}$ is scored on that basis. Because a departure is flown into the
wind wherever possible, and FAA vertiport guidance aligns preferred
approach/departure paths with the predominant wind to keep crosswind operations
to a minimum \citep{faa2024eb105}, the frequency with which wind blows from a
given azimuth is taken as the prevailing wind measure for a departure along that
azimuth, so that the azimuth and the reported wind direction share the same
reference.

Long term directional wind characteristics are evaluated using hourly surface
observations over a continuous ten year period ($W$). The ASOS record of the
Korea Meteorological Administration reports wind direction only in 16 compass
points (Sec.~\ref{sec:data}), so a finer division of the wind direction would
add no directional information. Observed directions are therefore aggregated
into 16 sectors of $22.5^{\circ}$ centered on those points, so that each sector
corresponds to exactly one reported direction and no sector is left empty by
the aggregation. Calm hours carry no direction and are excluded from the sector
frequencies. All azimuths within the same sector receive the same frequency
value and the same prevailing wind score. For each azimuth, the frequency of
winds from its sector is computed, and the prevailing wind score is defined as
follows.
\begin{equation}
S_{pw}(\theta) = \frac{f_{\theta}}{\max_{s \in \Theta} f_{s}}, \qquad
\Theta = \{\mathrm{N}, \mathrm{NNE}, \ldots, \mathrm{NNW}\}
\label{eq:spw}
\end{equation}
where $\Theta$ is the set of the 16 compass sectors, sector $s$
($s = 0, 1, \ldots, 15$) spanning $22.5^{\circ} s \pm 11.25^{\circ}$,
$f_{\theta}$ is the occurrence frequency of the sector containing azimuth
$\theta$, and $f_{s}$ that of sector $s$. This score expresses how often a
departure along that azimuth would be flown into a headwind, relative to the
most favorable azimuth.

\begin{figure}[htbp]
	\centering
		\includegraphics[width=0.68\linewidth]{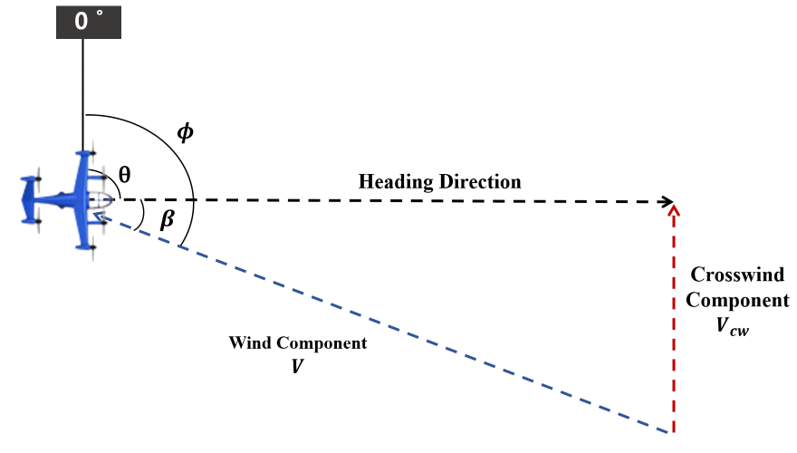}
	\caption{Crosswind component geometry for a candidate azimuth.}
	\label{fig:cwgeom}
\end{figure}

Crosswind operability is then evaluated for each candidate azimuth at each
observation time. The geometric relationship among the aircraft heading, wind
direction, angular difference, and crosswind component is shown in
Fig.~\ref{fig:cwgeom}. Observations are available at hourly intervals, each
time being denoted by $t$. The crosswind score is evaluated separately for
every $1^{\circ}$ azimuth. The angular difference between the observed wind
direction $\phi(t)$, taken as the direction code reported in the record (for
example $50^{\circ}$ for NE), and the candidate azimuth $\theta$ is
\begin{equation}
\beta(\theta,t) = \phi(t) - \theta
\label{eq:beta}
\end{equation}
and the corresponding crosswind component is computed as follows.
\begin{equation}
V_{cw}(\theta,t) = V(t) \times
\left| \sin\left( \tfrac{\pi}{180} \times \beta(\theta,t) \right) \right|
\label{eq:vcw}
\end{equation}
where $V(t)$ is the observed wind speed; for calm hours, which have no
direction, $V_{cw}$ is set to 0. Instantaneous crosswind operability is
quantified by the following penalty formulation.
\begin{equation}
S_{cw}(\theta,t,V_{cl}) =
\max\left( 0,\ 1 - \left( \frac{V_{cw}(\theta,t)}{V_{cl}} \right)^{q} \right)
\label{eq:scw}
\end{equation}
where $V_{cl} \in \{10.5, 13, 15\}$~kt denotes representative crosswind
limits. $V_{cl}$ is the reference against which each observation is scored: an
observation that exceeds it contributes 0 to the average, but an azimuth loses
its wind score entirely only if no observation falls below it. In the ten year
record used here the crosswind term therefore ranks azimuths rather than
excluding them. Multiple limits are considered to reflect differences in eVTOL
performance, and the three values are representative scenarios of crosswind
tolerance. The first two are the two lowest allowable crosswind components in
the FAA runway design criteria; because operational wind limits for eVTOLs have
not yet been standardized \citep{schweiger2023impact}, $15\,\mathrm{kt}$ was
evaluated in addition.

The parameter $q$ shapes how operability declines as the crosswind component
grows. For a normalized crosswind ratio between 0 and 1, a smaller $q$ returns a
lower score at every crosswind level and therefore gives a more conservative
evaluation, whereas a larger $q$ returns higher scores and preserves near full
operability until the crosswind approaches the limit. This study adopts
$q = 0.5$, a conservative choice that already lowers the score for crosswinds
well below the limit. Whatever the value of $q$, $S_{cw}$ reaches 0 when the
crosswind component reaches $V_{cl}$, so $q$ changes the magnitude of the wind
score and hence the ranking among usable azimuths rather than the set of usable
azimuths itself.

\begin{figure}[htbp]
	\centering
		\includegraphics[width=0.55\linewidth]{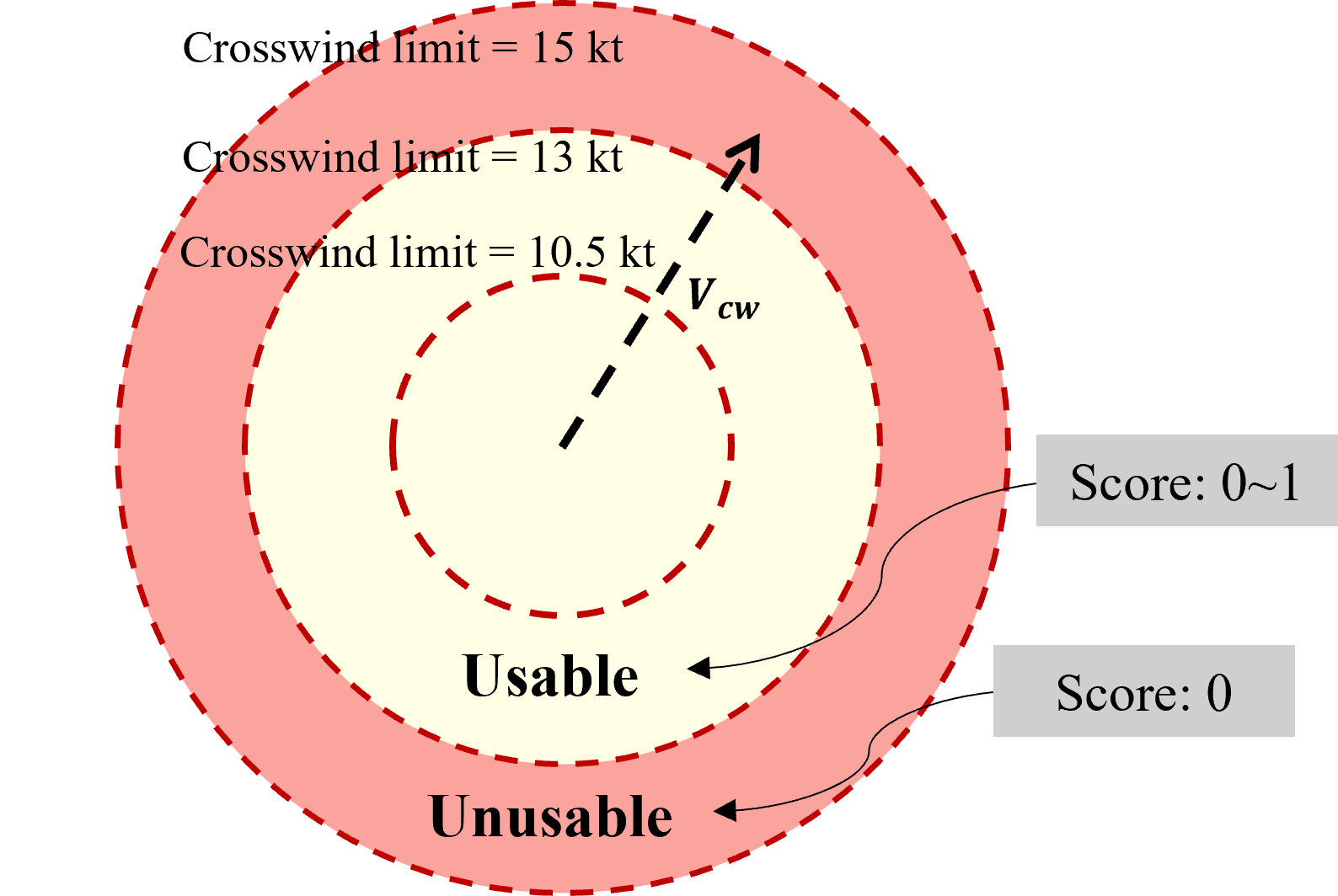}
	\caption{Crosswind penalty behavior under different allowable crosswind
limits.}
	\label{fig:cwpenalty}
\end{figure}

Figure~\ref{fig:cwpenalty} shows the penalty behavior for a case with a
crosswind component of $14\,\mathrm{kt}$. With the limit set to
$13\,\mathrm{kt}$, operability decreases continuously from 1 to 0 as the
crosswind approaches the limit and becomes 0 once the limit is exceeded. The
score produced under the same crosswind condition also differs according to
the selected limit.

For each azimuth, the instantaneous values are averaged over the ten year
period to obtain the long term mean crosswind operability.
\begin{equation}
S_{ac}(\theta,V_{cl}) = \frac{1}{|W|} \sum_{t \in W} S_{cw}(\theta,t,V_{cl})
\label{eq:sac}
\end{equation}
The wind operability score of each azimuth is defined as the product of the
prevailing wind score and the mean crosswind operability.
\begin{equation}
S_{wind}(\theta,V_{cl}) = S_{pw}(\theta) \times S_{ac}(\theta,V_{cl})
\label{eq:swind}
\end{equation}
The multiplicative form reflects the conjunctive nature of the two
requirements. An azimuth that is directionally unfavorable in the long term and
one that frequently exceeds the crosswind limit are both undesirable, and
neither compensates for the other. The complete procedure is summarized in
Algorithm~\ref{alg:wind}.

\begin{algorithm}[htbp]
\caption{Wind operability scoring by azimuth.}
\label{alg:wind}
\small
\begin{algorithmic}[1]
\Require ten year hourly wind record $W = \{(\phi(t),V(t))\}$;
  candidate azimuths $\theta \in \{0,1,\ldots,359\}^{\circ}$; crosswind limit $V_{cl}$
\Ensure wind operability score $S_{wind}(\theta,V_{cl})$
\Statex \emph{Step 1: long term prevailing wind analysis}
\ForAll{$s \in \Theta$}
  \State $f_{s} \gets |\{t \in W : \phi(t) \in s\}|$
  \Comment{$22.5^{\circ}$ sectors, calm excluded}
\EndFor
\ForAll{$\theta$}
  \State identify the sector containing $\theta$ and compute $S_{pw}(\theta)$
    by Eq.~\eqref{eq:spw}
\EndFor
\Statex \emph{Step 2: long term crosswind analysis}
\ForAll{$\theta$}
  \ForAll{$t \in W$}
    \State compute $S_{cw}(\theta,t,V_{cl})$ by Eqs.~\eqref{eq:beta}--\eqref{eq:scw}
  \EndFor
  \State compute $S_{ac}(\theta,V_{cl})$ by Eq.~\eqref{eq:sac}
\EndFor
\Statex \emph{Step 3: wind operability score by azimuth}
\ForAll{$\theta$}
  \State compute $S_{wind}(\theta,V_{cl})$ by Eq.~\eqref{eq:swind}
\EndFor
\State \Return $S_{wind}(\theta,V_{cl})$
\end{algorithmic}
\end{algorithm}

\section{Application to Two Vertiport Candidate Sites}
\label{sec:application}

\subsection{Case Study Areas}
\label{subsec:site}

\begin{figure}[htbp]
	\centering
		\includegraphics[width=0.7\linewidth]{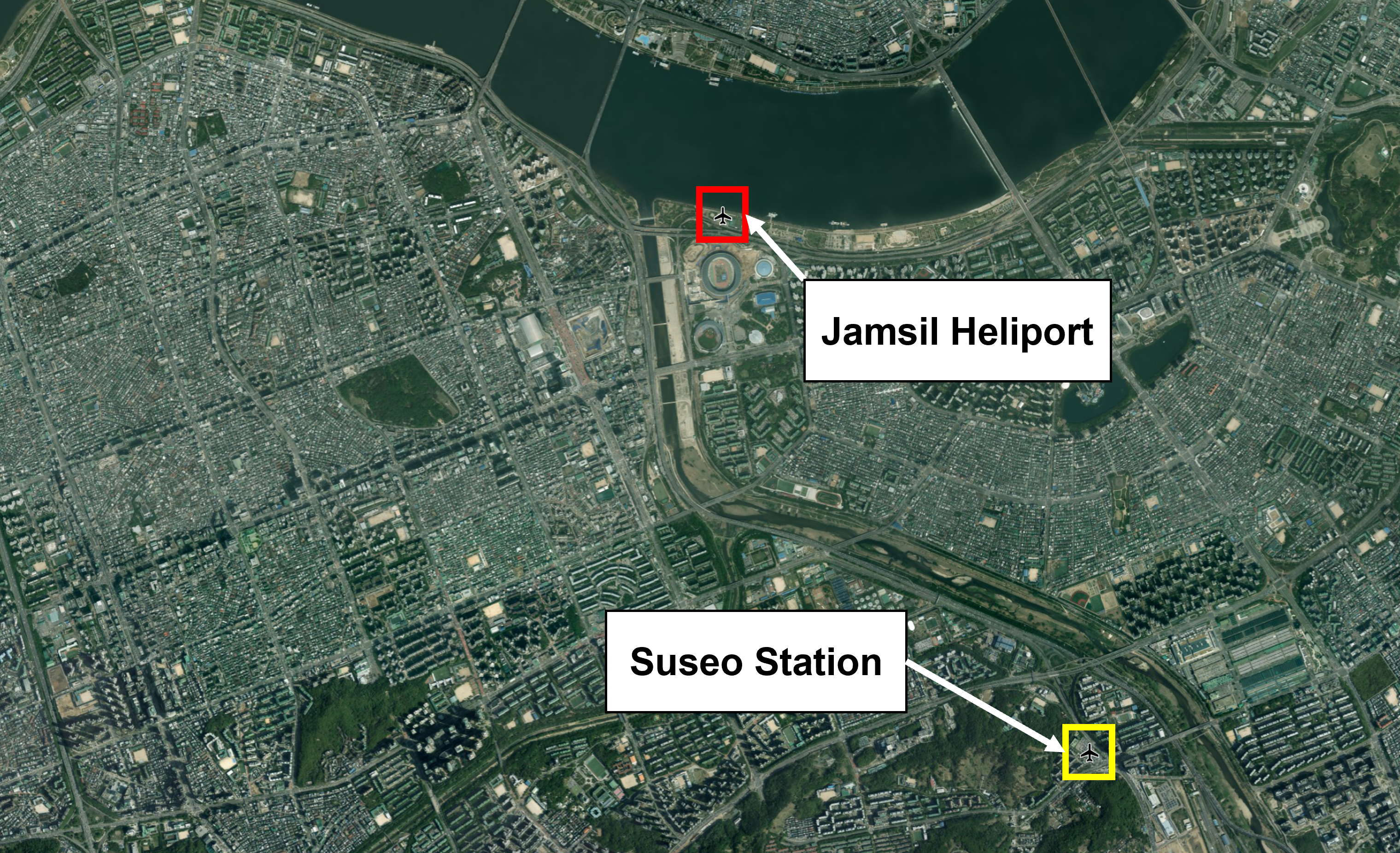}
	\caption{Locations of Suseo station (Case 1) and the Jamsil heliport (Case 2).}
	\label{fig:site}
\end{figure}

Two candidate sites in Seoul with contrasting surroundings, Suseo station and
the Jamsil heliport, are examined (Fig.~\ref{fig:site}). The two sites are the
two vertiports on the Tancheon route, the last of the initial routes of the
national Korean Urban Air Mobility (K-UAM) demonstration in the Seoul
metropolitan area and the one proposed to demonstrate operations entering the
urban core. The dedicated UAM aeronautical chart of the Ministry of Land,
Infrastructure and Transport presents this route together with the
surrounding obstacles and restricted airspace \citep{molit2023uamchart}.

Case 1 is Suseo station, an intermodal hub where the Super Rapid Train (SRT)
high speed rail, the Great Train eXpress Line A (GTX-A), subway Line 3, and the
Suin--Bundang Line meet. In some directions the surroundings have dense
clusters of tall residential buildings at close range, and in others
comparatively low development, so that obstacle conditions differ markedly with
direction.

Case 2 is the existing heliport at Jamsil on the Han River. Unlike Suseo
station, the site faces the river, so that most azimuths toward the north are
free of buildings, while the landward side is densely built.

The two cases show how directional obstacle and wind characteristics enter the
evaluation of departure directions at sites of different character.

\begin{figure}[p]
	\centering
	\subfloat[Suseo station]{%
	  \begin{minipage}{0.485\linewidth}\centering
	      \includegraphics[width=\linewidth,height=0.25\textheight,keepaspectratio]{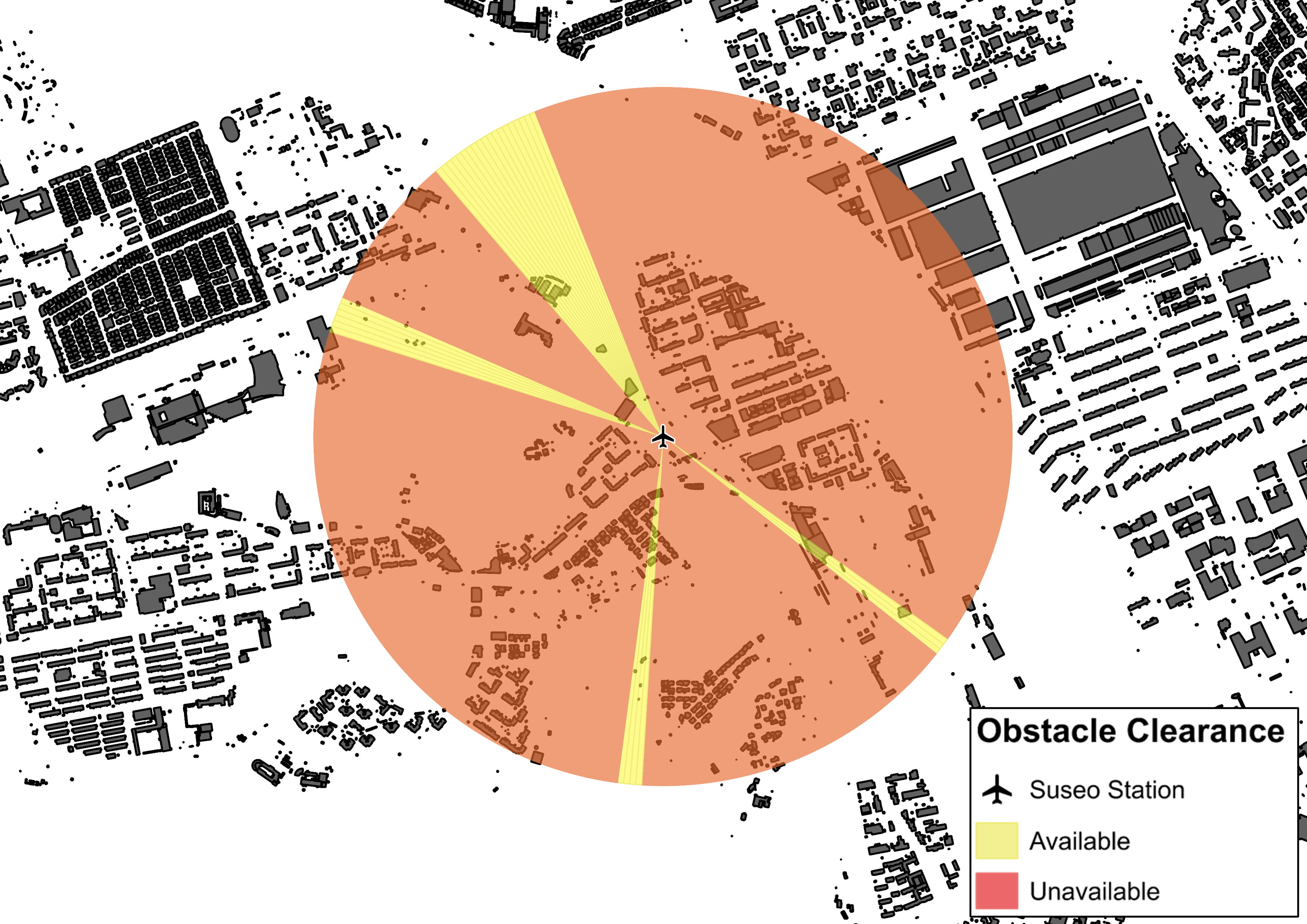}\\
	      {\small $5^{\circ}$ layer}\\[6pt]
	      \includegraphics[width=\linewidth,height=0.25\textheight,keepaspectratio]{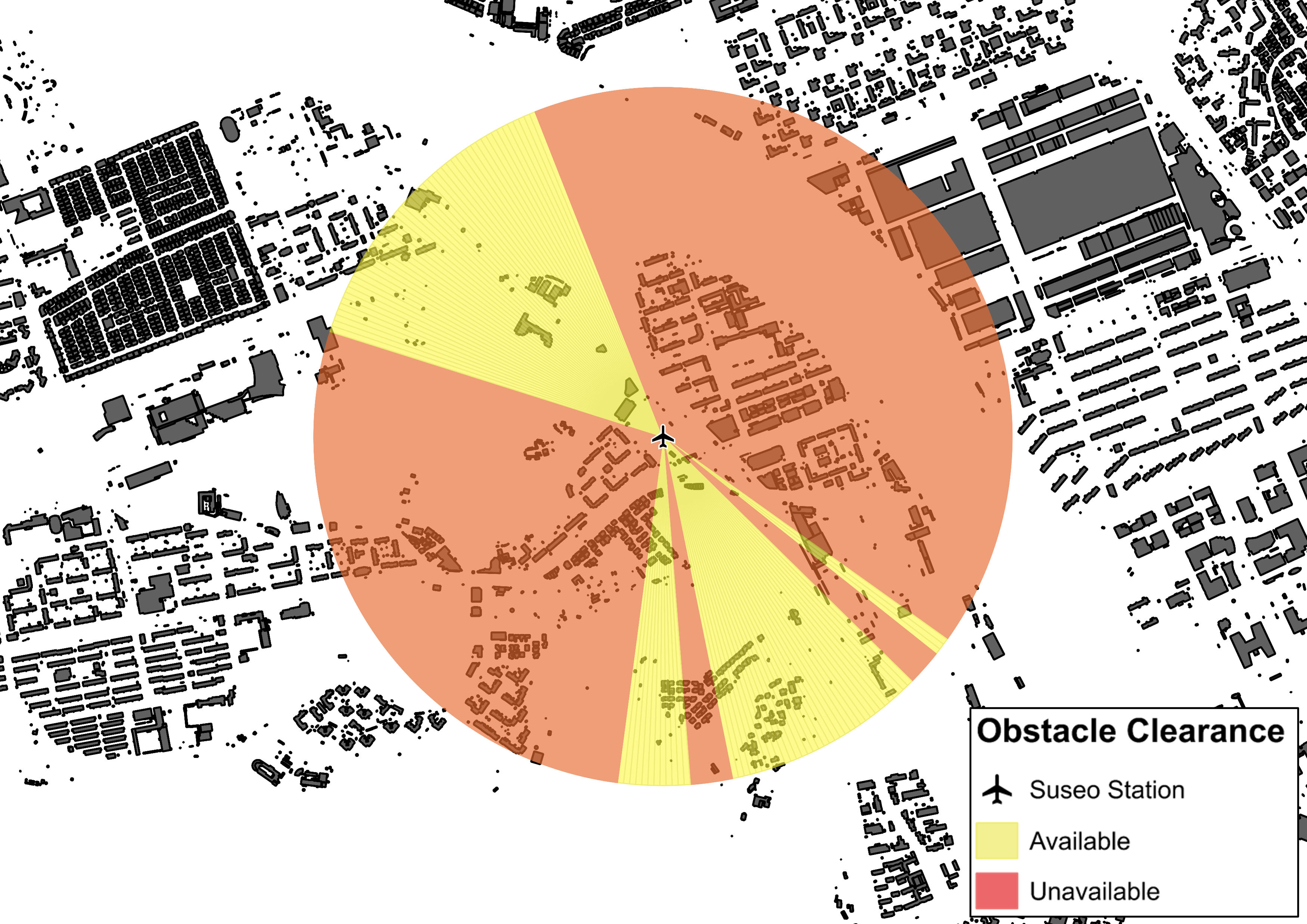}\\
	      {\small $8^{\circ}$ layer}\\[6pt]
	      \includegraphics[width=\linewidth,height=0.25\textheight,keepaspectratio]{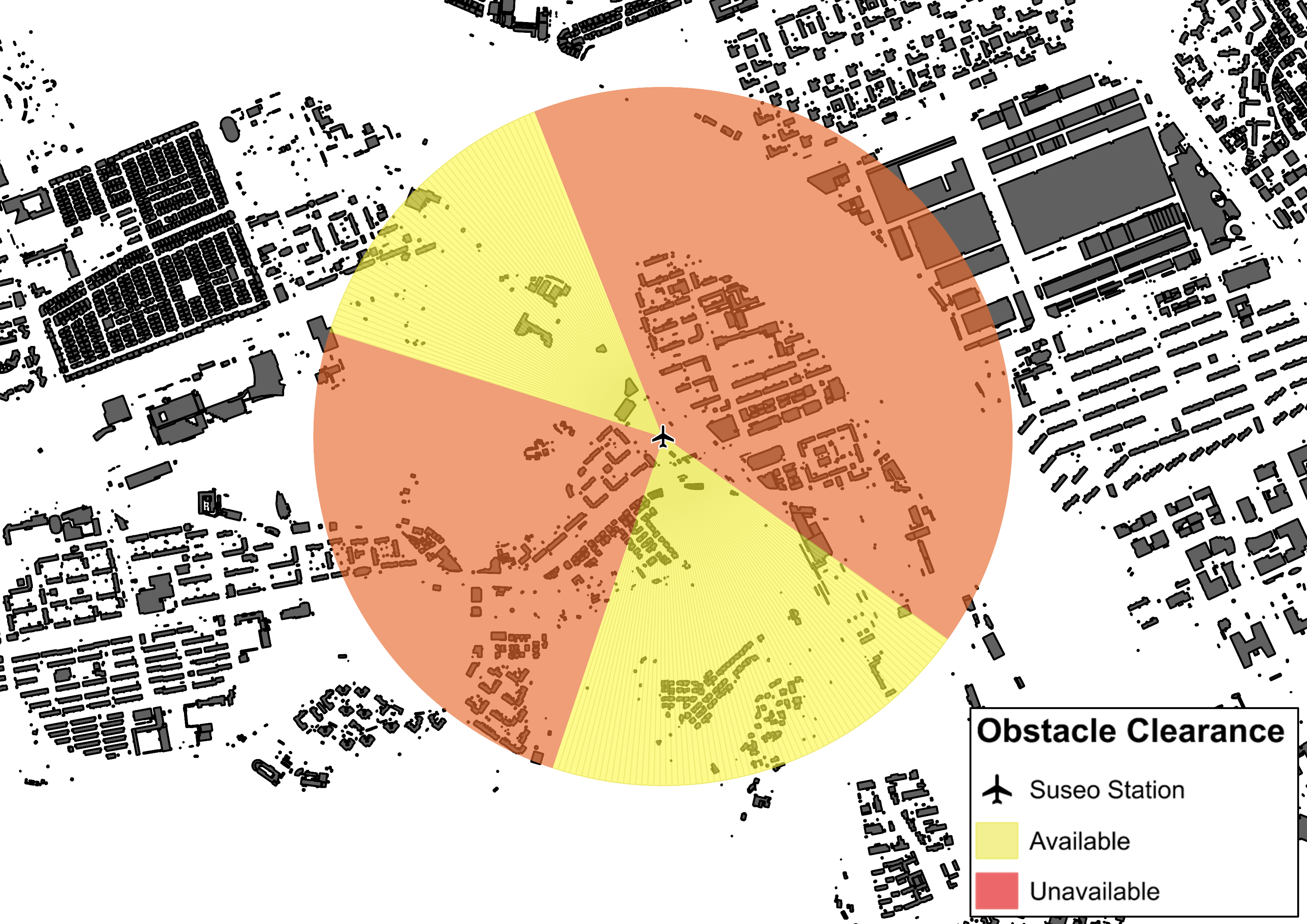}\\
	      {\small $12^{\circ}$ layer}
	  \end{minipage}}\hfill
	\subfloat[Jamsil heliport]{%
	  \begin{minipage}{0.485\linewidth}\centering
	      \includegraphics[width=\linewidth,height=0.25\textheight,keepaspectratio]{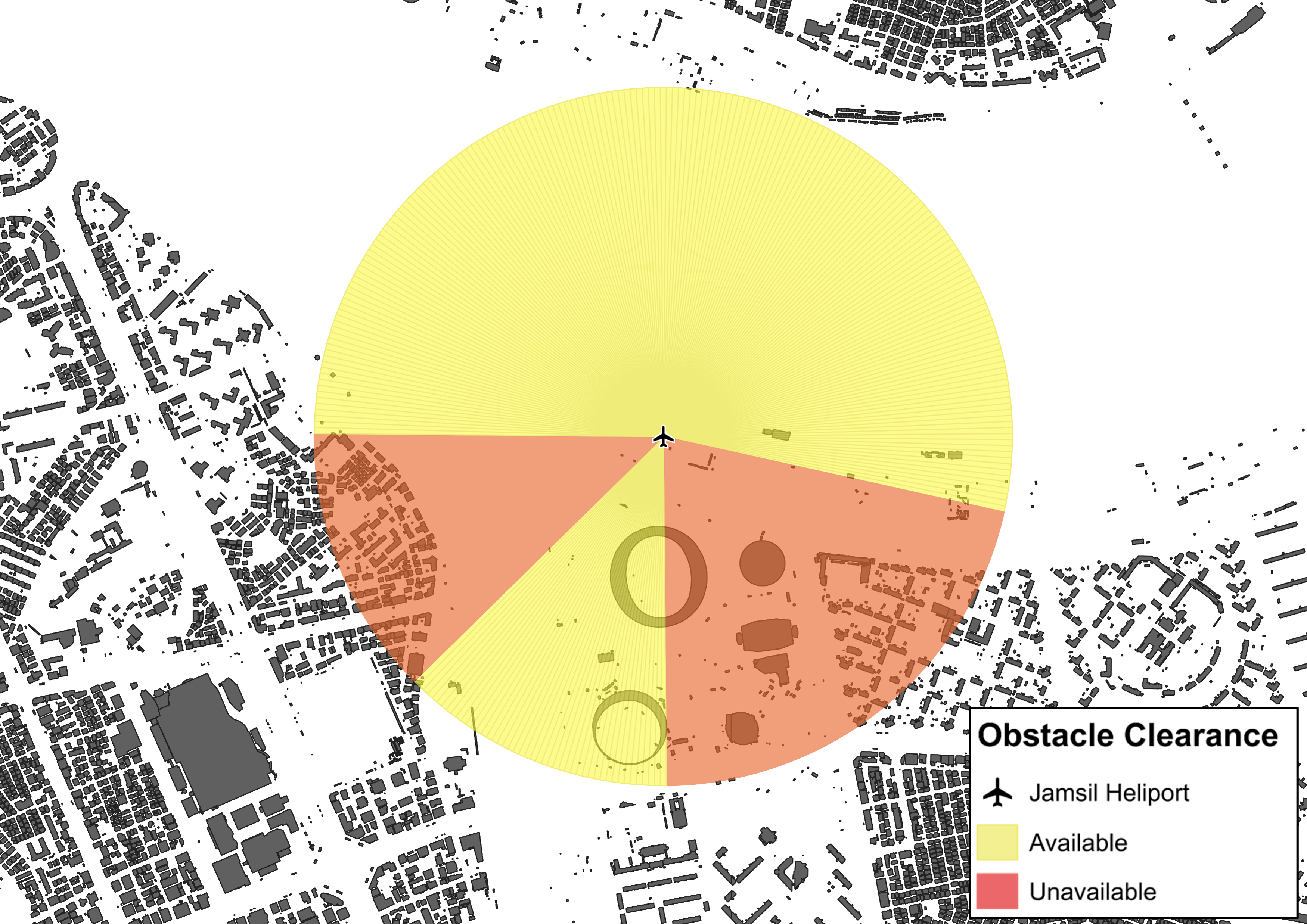}\\
	      {\small $5^{\circ}$ layer}\\[6pt]
	      \includegraphics[width=\linewidth,height=0.25\textheight,keepaspectratio]{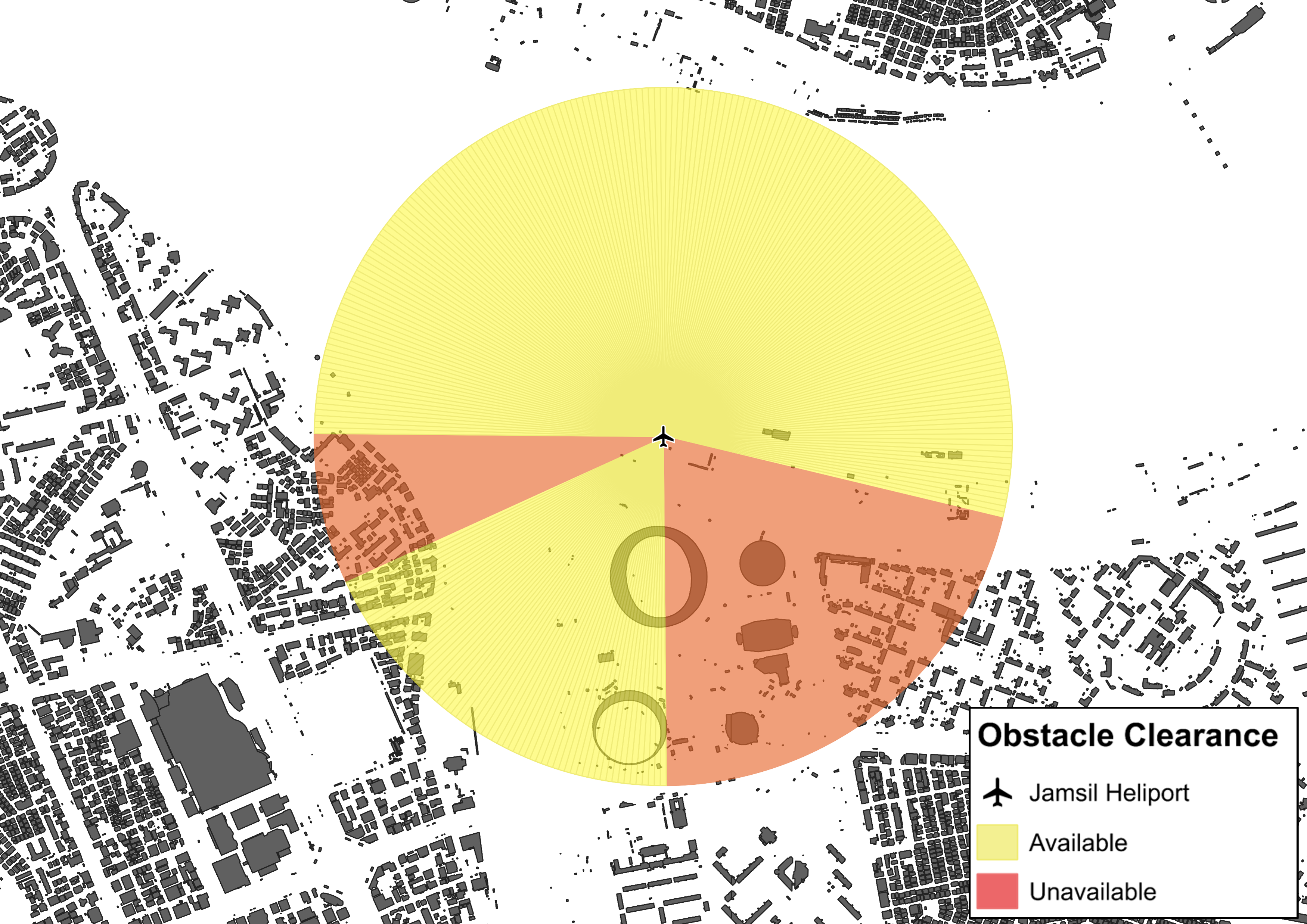}\\
	      {\small $8^{\circ}$ layer}\\[6pt]
	      \includegraphics[width=\linewidth,height=0.25\textheight,keepaspectratio]{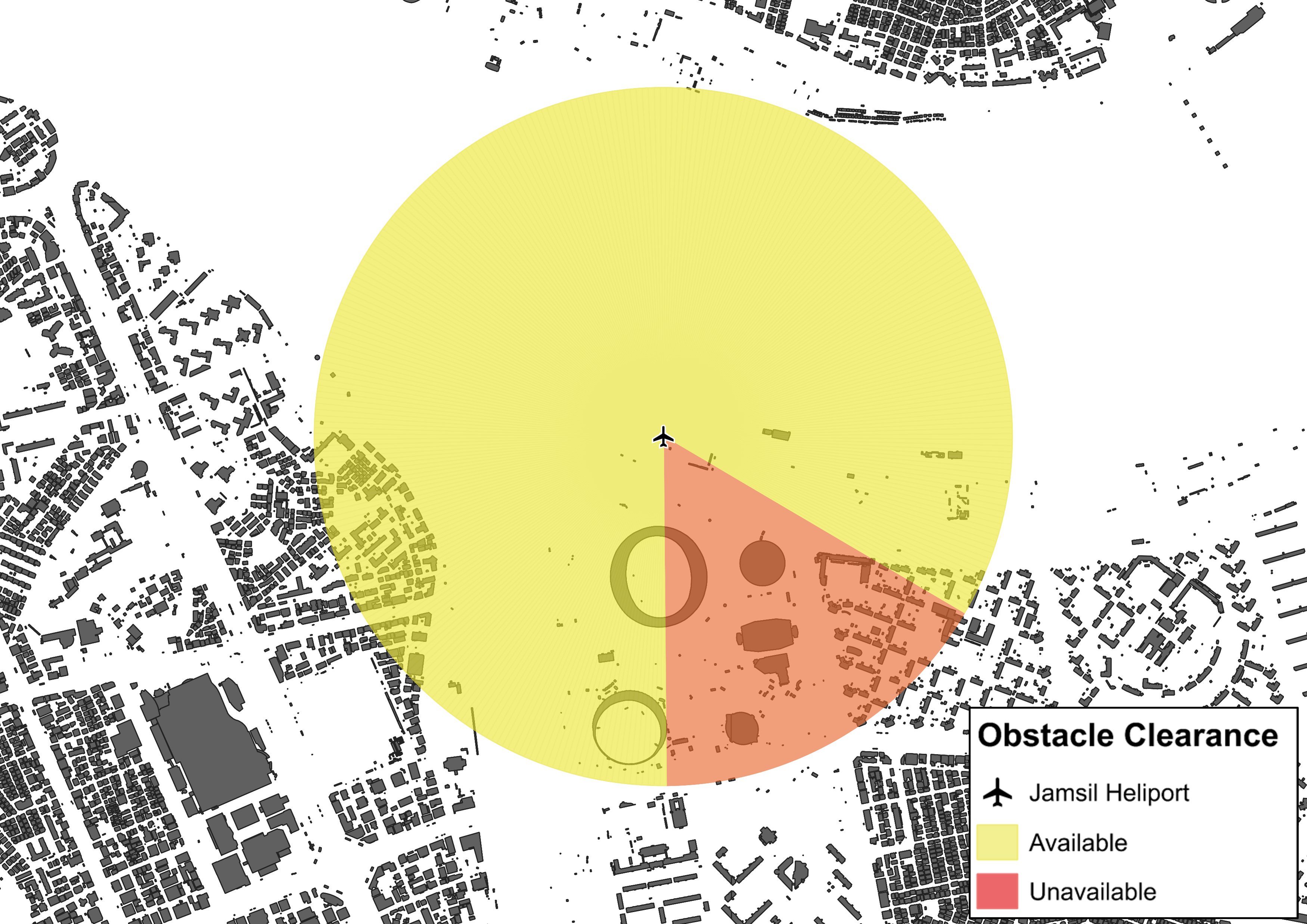}\\
	      {\small $12^{\circ}$ layer}
	  \end{minipage}}
	\caption{Usable azimuth sectors for each climb angle layer.}
	\label{fig:sectors}
\end{figure}

\subsection{Azimuth Evaluation by Obstacle Clearance}
\label{subsec:obsresult}

This section analyzes the results of applying the obstacle clearance scoring of
Sec.~\ref{subsec:combined} to the two sites. The analysis uses $p = 2$ as set in
Sec.~\ref{subsec:lateral}.

Figure~\ref{fig:sectors}a shows the azimuth sectors that satisfy
$S_{obs}(\theta,g) > 0$ at Case 1 in each climb angle layer. In the
$5^{\circ}$ layer the usable azimuth ranges were identified as
$126$--$128^{\circ}$, $184$--$187^{\circ}$, $288$--$293^{\circ}$, and
$320$--$338^{\circ}$, that is, 32 of the 360 evaluated azimuths, or about 8.9\%
of the azimuth space. Table~\ref{tab:obs} presents part of the obstacle level
evaluation results computed with the formulation of Sec.~\ref{sec:framework},
including both penetrating obstacles and obstacles with secured margins. The
method therefore reports the degree of margin or penetration for individual
buildings, going beyond a simple test of whether an obstacle penetrates. A
secured vertical margin at one obstacle does not by itself make its azimuth
usable, since the azimuth is excluded whenever any obstacle on it violates the
lateral corridor. In the $5^{\circ}$ layer a check along the centerline alone
would leave much wider ranges, and applying the width of the lateral corridor
reduces them to these 32 azimuths.

\begin{table}[htbp]
\caption{Obstacle level evaluation for the $5^{\circ}$ climb angle layer at Suseo station (excerpt). Heights are estimated at $4\,\mathrm{m}$ per story.}
\label{tab:obs}
\centering
\small
\begin{tabular}{lrrrrrrrr}
\toprule
Name & $H_{OBS,i}$ & Start angle & End angle & ${CM}_{i}$ & $d_{horizontal,i}$ &
$H_{OCS,i}$ & $H_{climb,i}$ & ${ROC}_{i}$ \\
 & (m) & ($^{\circ}$) & ($^{\circ}$) & (m) & (m) & (m) & (m) & (m) \\
\midrule
Suseo LH Apartments   & 60.00 & 0 & 13 & 45.25 & 213.78 & 14.75 & 24.80 & 10.05 \\
Samik Apartments      & 60.00 & 1 & 16 & 49.53 & 151.70 & 10.47 & 19.37 & 8.90 \\
Helio City Apartments & 140.00 & 16 & 19 & 80.49 & 862.55 & 59.51 & 81.56 & 22.05 \\
Hanareum Apartments   & 4.00 & 51 & 52 & $-25.02$ & 420.60 & 29.02 & 42.89 & 13.87 \\
Sungwoo Building      & 20.00 & 71 & 75 & $-8.23$ & 409.18 & 28.23 & 41.89 & 13.66 \\
\multicolumn{9}{c}{$\vdots$} \\
Suseo Apartments      & 60.00 & 305 & 309 & $-3.88$ & 925.90 & 63.88 & 87.10 & 23.22 \\
Suseo LH Apartments   & 4.00 & 355 & 357 & $-16.86$ & 302.33 & 20.86 & 32.55 & 11.69 \\
Suseo LH Apartments   & 60.00 & 359 & 0 & 29.16 & 446.97 & 30.84 & 45.20 & 14.36 \\
\bottomrule
\end{tabular}
\end{table}

\begin{figure}[htbp]
	\centering
		\subfloat[Suseo station]{\includegraphics[width=\linewidth]{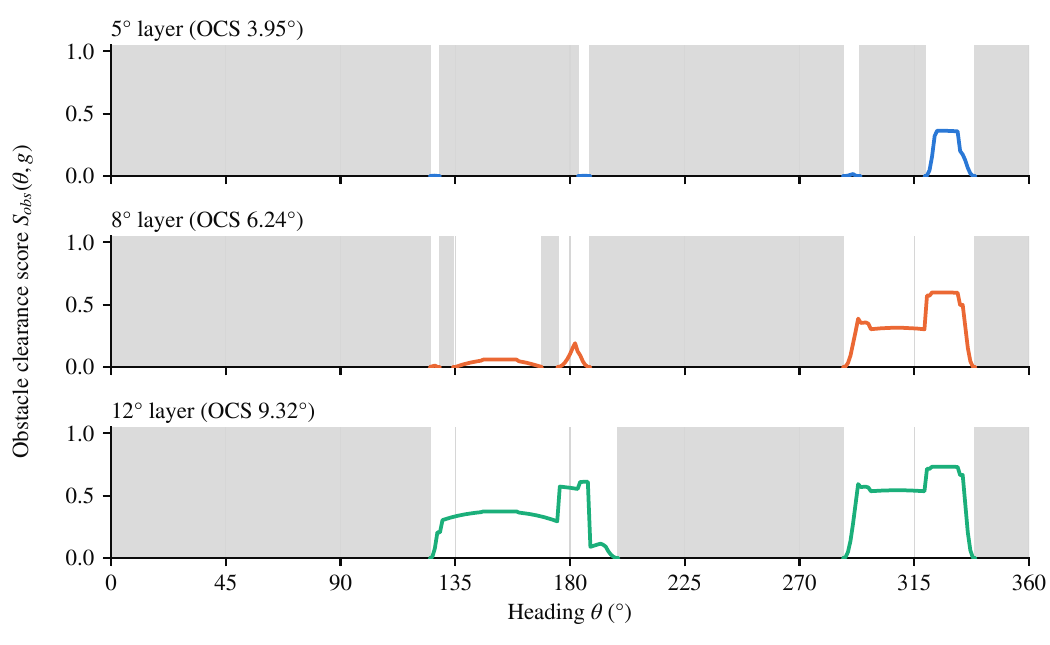}}\\[2pt]
		\subfloat[Jamsil heliport]{\includegraphics[width=\linewidth]{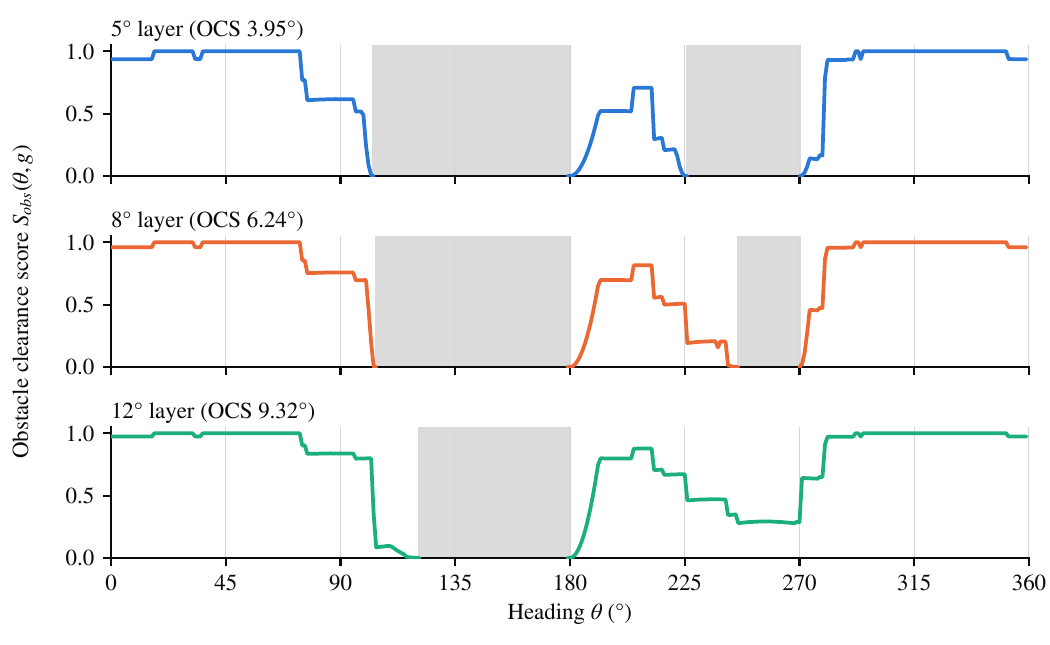}}
	\caption{Obstacle clearance score $S_{obs}(\theta,g)$ by climb angle layer; shaded bands mark unusable azimuths.}
	\label{fig:layercmp}
\end{figure}

Figure~\ref{fig:layercmp} compares the obstacle clearance score distributions of
the three layers at both sites. For Case 1, in the $8^{\circ}$ layer the usable
azimuth range expands to $126$--$128^{\circ}$, $135$--$168^{\circ}$,
$176$--$187^{\circ}$, and $288$--$338^{\circ}$, 100 azimuths in total
(27.8\%). In the $12^{\circ}$ layer it expands further to $126$--$198^{\circ}$
and $288$--$338^{\circ}$, that is, 124 azimuths or 34.4\%.

To compare the layers quantitatively, directional availability is defined as
the share of the 360 evaluated azimuths that are usable.
\begin{equation}
A_{g} = \frac{N_{g}^{usable}}{360}
\label{eq:availability}
\end{equation}
where $N_{g}^{usable}$ is the number of usable azimuths in layer $g$. Each
azimuth is evaluated at $1^{\circ}$ intervals and stands for a $1^{\circ}$
sector, so $A_{g}$ is the share of the azimuth domain that is usable. By this
measure, $A_{5} = 0.089$, $A_{8} = 0.278$, and $A_{12} = 0.344$. Even with the
same vertiport facility and the same obstacle environment, the number of usable
azimuths varies with the allowable climb profile, so that vehicle capability,
and not facility conditions alone, determines part of the directional resource
available. Directional availability here is determined by the obstacle
condition alone.

Most of the gain from the $5^{\circ}$ to the $8^{\circ}$ layer comes from a few
obstacles. Two low buildings about $37$ and $55\,\mathrm{m}$ from the FATO edge
lie inside the core corridor where the $5^{\circ}$ protection surface is only a
few meters high and close $135$--$168^{\circ}$ and $176$--$183^{\circ}$, and a
high-rise building about $850\,\mathrm{m}$ along the path closes most of
$294$--$319^{\circ}$; all three sectors reopen in the $8^{\circ}$ layer, so
aircraft capable of a steeper climb gain directions unavailable to aircraft
limited to a shallow climb. The nearer of the two buildings and a third low
building keep $129$--$134^{\circ}$, $169$--$175^{\circ}$, and
$188$--$198^{\circ}$ closed until the $12^{\circ}$ layer, so near the FATO,
where the protection surface is lowest, a few low buildings can close a wide
sector for a shallow climb layer. Beyond $8^{\circ}$ the gain is smaller: the
24 added azimuths all lie in this southern sector, while $199$--$287^{\circ}$
and $339$--$125^{\circ}$ remain closed in every layer because their obstacles
reach even the steepest protection surface. Reopening a sector also does not
give it a wide margin: $S_{obs}$ in $126$--$198^{\circ}$ stays at or below
0.19 in the $8^{\circ}$ layer and 0.61 in the $12^{\circ}$ layer, against 0.60
and 0.73 in the northwestern sector (Fig.~\ref{fig:layercmp}a). Because the
$8^{\circ}$ surface clears the nearest low building, estimated at
$4\,\mathrm{m}$, by at most about $0.25\,\mathrm{m}$, the opening of
$135$--$168^{\circ}$ in that layer, and with it $A_{8}$, is sensitive to the
height estimated from the number of stories, as are azimuths with a near zero
$S_{obs}$, such as $290^{\circ}$ in the $5^{\circ}$ layer.

Case 2 presents a different obstacle environment (Figs.~\ref{fig:sectors}b and
\ref{fig:layercmp}b). In the $5^{\circ}$ layer the usable ranges are
$180$--$225^{\circ}$ and $271$--$102^{\circ}$, 238 azimuths in total
($A_{5} = 0.661$); they widen to 259 ($A_{8} = 0.719$) in the $8^{\circ}$ layer
and to 301 ($A_{12} = 0.836$) in the $12^{\circ}$ layer, where only
$121$--$179^{\circ}$ remains closed. $S_{obs} = 1$ in every layer over
$17$--$32^{\circ}$, $36$--$74^{\circ}$, $292$--$293^{\circ}$, and
$295$--$351^{\circ}$, most of which face the river. The sector that stays closed
in every layer is governed by a single building about $20\,\mathrm{m}$ from the
FATO edge, whereas the sectors that reopen in the steeper layers,
$226$--$270^{\circ}$ and $103$--$120^{\circ}$, are closed in the $5^{\circ}$
layer mainly by tall buildings about $670$--$900\,\mathrm{m}$ along the path.
The layer effect is therefore present at both sites, but its size and its source
depend on the surroundings: at Case 1 the $5^{\circ}$ layer keeps only 8.9\% of
the azimuth space and gains most from a steeper climb, whereas at Case 2 it
already keeps about two thirds (66.1\%), and in the steeper layers the sectors
blocked by distant tall buildings become usable.

\subsection{Azimuth Evaluation by Wind Operability}
\label{subsec:windresult}

The scoring of Sec.~\ref{sec:wind} was applied to the ten year ASOS
record \citep{kma2025} described in Sec.~\ref{sec:data}. Because both sites use
the same record, the wind scores below apply to both. The prevailing wind
analysis showed the highest occurrence frequency in the W sector centered on
$270^{\circ}$, followed by the WNW and NE sectors, and Table~\ref{tab:wind}
summarizes the observed frequencies and the corresponding normalized scores.
Because the sectors follow the 16 compass points of the record, every sector
contains observations, and even the lowest scores, in the SE and SSE sectors,
are above zero.

\begin{table}[htbp]
\caption{Prevailing wind frequency and normalized score by compass sector. Each
sector is $22.5^{\circ}$ wide and centered on its compass point; calm hours are
excluded.}
\label{tab:wind}
\centering
\small
\begin{tabular}{cccc@{\hspace{24pt}}cccc}
\toprule
Sector & Center & $f_{s}$ & $S_{pw}$ & Sector & Center & $f_{s}$ & $S_{pw}$ \\
 & ($^{\circ}$) & (h) & & & ($^{\circ}$) & (h) & \\
\midrule
N   & 0     & 2394  & 0.173 & S   & 180   & 2137  & 0.154 \\
NNE & 22.5  & 5068  & 0.366 & SSW & 202.5 & 4606  & 0.333 \\
NE  & 45    & 10832 & 0.783 & SW  & 225   & 4544  & 0.328 \\
ENE & 67.5  & 7287  & 0.527 & WSW & 247.5 & 7267  & 0.525 \\
E   & 90    & 3021  & 0.218 & W   & 270   & 13835 & 1.000 \\
ESE & 112.5 & 1957  & 0.141 & WNW & 292.5 & 11507 & 0.832 \\
SE  & 135   & 1340  & 0.097 & NW  & 315   & 5107  & 0.369 \\
SSE & 157.5 & 1216  & 0.088 & NNW & 337.5 & 2586  & 0.187 \\
\bottomrule
\end{tabular}
\end{table}

Crosswind scores were then computed with $q = 0.5$ following the formulation of
Sec.~\ref{sec:wind} and combined with the prevailing wind scores to obtain the
wind operability score $S_{wind}(\theta,V_{cl})$ for each azimuth. Across the
three crosswind limits the distributions were similar in shape, since
$S_{pw}(\theta)$ is identical in all three cases and the crosswind decay terms
differ little. In the record used here, the crosswind component on the leading
candidate azimuths exceeded $10.5\,\mathrm{kt}$ in less than 0.6\% of hours,
$13\,\mathrm{kt}$ in less than 0.1\%, and $15\,\mathrm{kt}$ in about 0.01\%, so
the choice of limit has little effect on the ranking, and the results that
follow are reported for $V_{cl} = 10.5\,\mathrm{kt}$ alone. At a site with stronger crosswinds
the three limits would separate, and the sensitivity to $V_{cl}$ and to $q$
would have to be re-examined and the parameters adjusted accordingly. The mean
crosswind operability was also nonzero for every azimuth, so $S_{wind}$ is
positive at every azimuth and no azimuth is excluded on wind grounds. The wind
score serves to identify directions that are unfavorable over the long term and
to capture the general pattern of directional operability.

\subsection{Integrated Score and Weighting Scenario Analysis}
\label{subsec:weighting}

The obstacle clearance score and the wind operability score are combined into
an integrated score for each combination of azimuth $\theta$, climb angle layer
$g$, and crosswind limit $V_{cl}$. The weights used in the combination are drawn
from the set $K$ of weighting scenarios listed in Table~\ref{tab:weights}, and
in what follows $k \in K$ denotes one of these scenarios.
\begin{equation}
S_{total}^{(k)}(\theta,g,V_{cl}) = w_{obs}^{(k)} S_{obs}(\theta,g)
+ w_{wind}^{(k)} S_{wind}(\theta,V_{cl})
\label{eq:stotal}
\end{equation}
where $w_{obs}^{(k)}$ and $w_{wind}^{(k)}$ are the obstacle and wind weights of
scenario $k$, with $w_{obs}^{(k)} + w_{wind}^{(k)} = 1$ for every scenario. A
strict usability condition is imposed before the weights are applied: if either
component is 0 for a given azimuth, the integrated score is set to 0. An azimuth
with no obstacle clearance margin, or with a wind score of 0, is thus excluded
regardless of the magnitude of the other component, and the weights affect only
the ordering of the usable candidates. In the record used here $S_{wind}$ is
positive at every azimuth (Sec.~\ref{subsec:windresult}), so the usable set is
determined by the obstacle condition alone and the wind score expresses an
operational preference among the usable azimuths.

The candidate space consists of combinations of 360 azimuths and three climb
angle layers, and the total number of evaluations remains limited even when the
crosswind limits and weighting scenarios are included. This study therefore
evaluates all combinations exhaustively.

No quantitative standard currently prescribes the relative importance of the
obstacle clearance margin and wind based operability in evaluating vertiport
departure directions. Four representative weighting scenarios are therefore
applied, as listed in Table~\ref{tab:weights}, in place of a single definitive
weight combination. They serve to analyze how different priority settings
affect the ranking and selection among directions that are already usable.

\begin{table}[htbp]
\caption{Weighting scenario configuration.}
\label{tab:weights}
\centering
\small
\begin{tabular}{lcc}
\toprule
Scenario & $w_{obs}$ & $w_{wind}$ \\
\midrule
Obstacle dominant    & 0.9 & 0.1 \\
Obstacle prioritized & 0.7 & 0.3 \\
Balanced             & 0.5 & 0.5 \\
Wind prioritized     & 0.3 & 0.7 \\
\bottomrule
\end{tabular}
\end{table}

Table~\ref{tab:optimal} lists the highest scoring azimuth and its integrated
score for each weighting scenario and climb angle layer at both sites under a
crosswind limit of $V_{cl} = 10.5\,\mathrm{kt}$.

\begin{table}[htbp]
\caption{Highest scoring azimuth $\theta$ and integrated score by weighting scenario and climb angle layer ($V_{cl} = 10.5\,\mathrm{kt}$).}
\label{tab:optimal}
\centering
\small
\begin{tabular}{cccccc}
\toprule
 & & \multicolumn{2}{c}{Suseo station} & \multicolumn{2}{c}{Jamsil heliport} \\
\cmidrule(lr){3-4}\cmidrule(lr){5-6}
Layer & $w_{obs} : w_{wind}$ & $\theta$ ($^{\circ}$) & Score & $\theta$ ($^{\circ}$) & Score \\
\midrule
\multirow{4}{*}{$5^{\circ}$}
 & 0.9 : 0.1 & 324 & 0.3448 & 50 & 0.9470 \\
 & 0.7 : 0.3 & 324 & 0.3081 & 50 & 0.8410 \\
 & 0.5 : 0.5 & 324 & 0.2715 & 281 & 0.7525 \\
 & 0.3 : 0.7 & 290 & 0.3443 & 281 & 0.6812 \\
\midrule
\multirow{4}{*}{$8^{\circ}$}
 & 0.9 : 0.1 & 322 & 0.5564 & 50 & 0.9470 \\
 & 0.7 : 0.3 & 322 & 0.4732 & 281 & 0.8417 \\
 & 0.5 : 0.5 & 293 & 0.4255 & 281 & 0.7653 \\
 & 0.3 : 0.7 & 293 & 0.4405 & 281 & 0.6889 \\
\midrule
\multirow{4}{*}{$12^{\circ}$}
 & 0.9 : 0.1 & 322 & 0.6771 & 50 & 0.9470 \\
 & 0.7 : 0.3 & 322 & 0.5670 & 281 & 0.8519 \\
 & 0.5 : 0.5 & 293 & 0.5276 & 281 & 0.7726 \\
 & 0.3 : 0.7 & 293 & 0.5017 & 281 & 0.6932 \\
\bottomrule
\end{tabular}
\end{table}

\begin{figure}[htbp]
	\centering
		\subfloat[Suseo station]{\includegraphics[width=\linewidth]{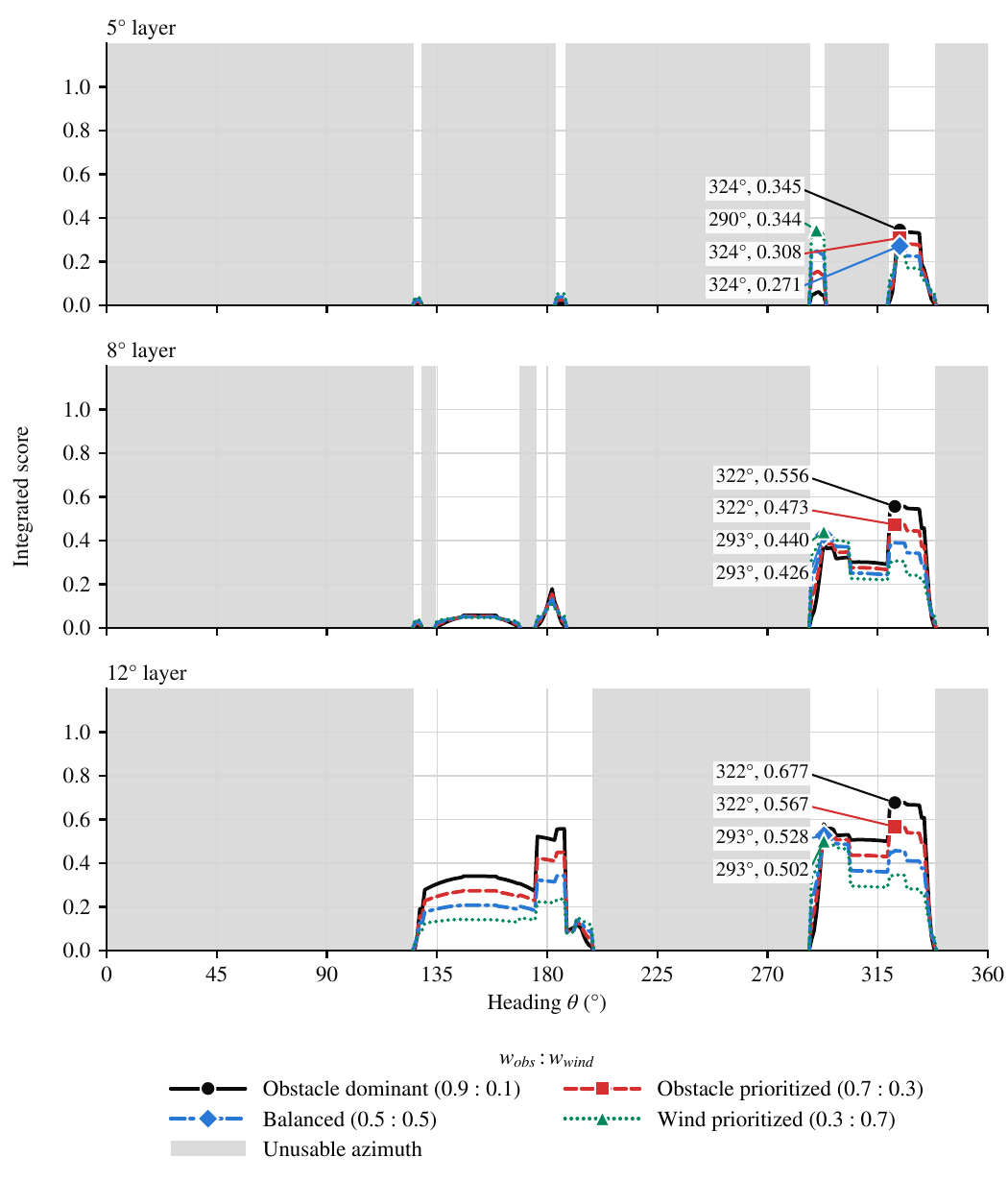}}
	\caption{Integrated score by azimuth for the four weighting scenarios ($V_{cl} = 10.5\,\mathrm{kt}$).}
	\label{fig:total}
\end{figure}

\begin{figure}[htbp]
	\ContinuedFloat
	\centering
		\subfloat[Jamsil heliport]{\includegraphics[width=\linewidth]{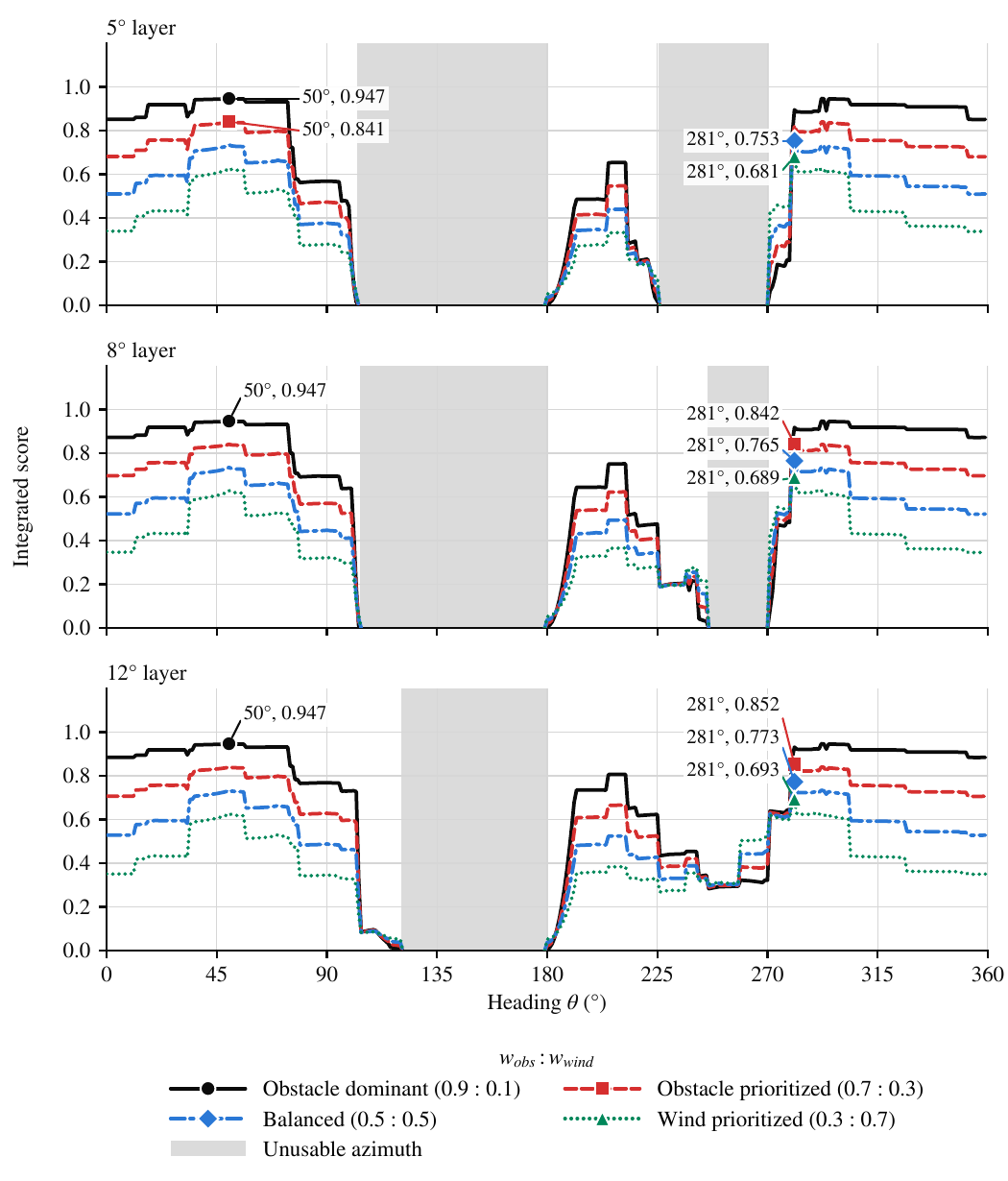}}
	\caption{Integrated score by azimuth for the four weighting scenarios ($V_{cl} = 10.5\,\mathrm{kt}$) (continued).}
\end{figure}

Figure~\ref{fig:total} shows the same results over the entire azimuth domain.
The shaded bands mark the azimuths excluded by the usability condition and are
identical across the four scenarios, showing that the weights operate only
within the unshaded ranges. Within those ranges the four curves differ in
shape, because the two components vary differently across azimuths, and the
filled markers show where the highest scoring azimuth of each scenario lies.

Two results stand out. First, at both sites the directions with the largest
obstacle margin differ from those facing the prevailing wind, and the two
groups trade off against each other. At Case 1, when obstacle conditions are
weighted heavily, the leading candidate lies in the northwestern sector, at
$324^{\circ}$ in the $5^{\circ}$ layer and $322^{\circ}$ in the $8^{\circ}$
and $12^{\circ}$ layers. These azimuths have the largest obstacle margin in
their layers, with $S_{obs}$ of 0.36, 0.60, and 0.73 set by a building about
$270\,\mathrm{m}$ along the path, but face an infrequent wind sector
($S_{wind}$ of 0.18). As the weight on wind increases, the leading candidate
shifts to $290^{\circ}$ in the $5^{\circ}$ layer and to $293^{\circ}$ in the
$8^{\circ}$ and $12^{\circ}$ layers, which face the prevailing westerly winds
($S_{wind}$ of 0.46 to 0.49) but have a smaller obstacle margin. At Case 2 the
obstacle favored direction is $50^{\circ}$, which is free of obstacles in every
layer ($S_{obs} = 1$) and faces the NE sector, the third most frequent
($S_{wind} = 0.47$); the equally unobstructed $292^{\circ}$ is virtually tied
with it ($S_{wind} = 0.468$). The wind favored direction is $281^{\circ}$, at the edge of
the most frequent W sector ($S_{wind} = 0.57$), where buildings about
$840\,\mathrm{m}$ along the path leave $S_{obs}$ of 0.93 to 0.97. The azimuths
nearer the center of the W sector, $271$--$279^{\circ}$, are penalized more
strongly by buildings about $740$--$770\,\mathrm{m}$ along the path and do not
lead. In the linear combination of Eq.~\eqref{eq:stotal}, the obstacle favored
directions dominate when $w_{obs}$ is large and the wind favored directions
when $w_{wind}$ is large.

Second, the point at which the ranking reverses differs across climb angle
layers at both sites. Solving Eq.~\eqref{eq:stotal} for the weight at which the
obstacle favored and the wind favored azimuth of each layer score equally gives
$w_{wind} \approx 0.534$, $0.428$, and $0.332$ for the $5^{\circ}$,
$8^{\circ}$, and $12^{\circ}$ layers at Case 1 and $0.399$, $0.295$, and
$0.218$ at Case 2. The reversal occurs at a lower wind weight as the climb angle
increases because the wind favored direction retains a larger obstacle margin
in the steeper layers: at Case 1, $S_{obs}$ at the wind favored azimuth is 0.01
at $290^{\circ}$ in the $5^{\circ}$ layer and 0.39 and 0.59 at $293^{\circ}$ in
the $8^{\circ}$ and $12^{\circ}$ layers, and at Case 2 $S_{obs}$ at
$281^{\circ}$ rises from 0.93 to 0.97. At Case 1 the leading candidate therefore
moves to the wind favored azimuth in the balanced scenario in the $8^{\circ}$
and $12^{\circ}$ layers but only in the wind prioritized scenario in the
$5^{\circ}$ layer. At Case 2 it moves in the balanced scenario in the
$5^{\circ}$ layer and already in the obstacle prioritized scenario in the
$8^{\circ}$ and $12^{\circ}$ layers, whose reversal weights of 0.295 and 0.218
lie below that scenario's wind weight of 0.3, in the $8^{\circ}$ layer by a
narrow margin. Even under the same operational priority, the preferred
departure direction thus differs according to the climb angle layer available
to the vehicle; in the balanced scenario at Case 1, for example, the
$5^{\circ}$ layer selects $324^{\circ}$ while the $8^{\circ}$ and $12^{\circ}$
layers select $293^{\circ}$. The two components also span different ranges
among usable azimuths at Case 1, $S_{obs}$ reaching 0.36 to 0.73 at the
obstacle favored leading azimuths while $S_{wind}$ does not exceed about 0.49,
so equal nominal weights do not give them equal influence on the ranking.

Because the wind record resolves direction only to 16 compass points, the
ranking among individual azimuths within a compass sector partly reflects that
resolution. $S_{ac}$ peaks at the reported direction codes, such as $50^{\circ}$
and $290^{\circ}$, and $S_{pw}$ changes in steps at the sector boundaries, such
as $281^{\circ}$ and $56^{\circ}$. Each leading azimuth is therefore read as
representative of its sector.

These results show that the preferred departure direction is determined jointly
by operational priorities, the climb profile available to the aircraft, and the
arrangement of obstacles and prevailing winds around the site.

Applying the usability condition of Eq.~\eqref{eq:stotal} yields, for each
weighting scenario $k$, one candidate set $P_{g,V_{cl}}^{(k)}$ for each of the
nine combinations of climb angle layer and crosswind limit, containing the
usable azimuths of that layer ranked by integrated score. Because $S_{obs}$ is
normalized within each layer, the ranking is made within a layer and not across
layers. The complete procedure is summarized in Algorithm~\ref{alg:rank}.

\begin{algorithm}[htbp]
\caption{Candidate ranking by weighting scenario.}
\label{alg:rank}
\small
\begin{algorithmic}[1]
\Require $S_{obs}(\theta,g)$ from Algorithm~\ref{alg:obs};
  $S_{wind}(\theta,V_{cl})$ from Algorithm~\ref{alg:wind};
  set of weighting scenarios $K$ (Table~\ref{tab:weights})
\Ensure ranked candidate sets $P_{g,V_{cl}}^{(k)}$
\ForAll{$k = ( w_{obs}^{(k)}, w_{wind}^{(k)} ) \in K$}
  \ForAll{$V_{cl} \in \{10.5, 13, 15\}$~kt}
    \ForAll{$g$}
      \State $P_{g,V_{cl}}^{(k)} \gets \varnothing$
      \ForAll{$\theta$}
        \If{$S_{obs}(\theta,g) = 0$ \textbf{or} $S_{wind}(\theta,V_{cl}) = 0$}
          \State \textbf{continue}
          \Comment{unusable azimuth}
        \EndIf
        \State compute $S_{total}^{(k)}(\theta,g,V_{cl})$ by Eq.~\eqref{eq:stotal}
        \State $P_{g,V_{cl}}^{(k)} \gets P_{g,V_{cl}}^{(k)} \cup
          \{ ( \theta, S_{total}^{(k)} ) \}$
      \EndFor
      \State sort $P_{g,V_{cl}}^{(k)}$ by $S_{total}^{(k)}$ in descending order
    \EndFor
  \EndFor
\EndFor
\State \Return $\{ P_{g,V_{cl}}^{(k)} \}$
\end{algorithmic}
\end{algorithm}

\section{Selection of Directionally Separated Candidates}
\label{sec:reselection}

The proposed framework produces a ranked set of usable directions rather than a
single direction. This section illustrates how such a set is reduced to a small
number of directionally separated candidates. The framework does not generate
new directions in real time, which would raise certification difficulties. It
identifies in advance the directions that satisfy the obstacle clearance
requirement, ranks them by the combined obstacle and wind score, and confines
any later reconfiguration to that pre-screened candidate set.

The selection reported here uses the obstacle prioritized scenario
($w_{obs} : w_{wind} = 0.7 : 0.3$) and a crosswind limit of
$10.5\,\mathrm{kt}$. Within each climb angle layer, candidates are drawn from
the ranked set in descending order of $S_{total}$ subject to two conditions.
The integrated score must reach a minimum admissible value $\tau = \taumin$,
and any two selected directions must be at least $135^{\circ}$ apart. This
value is the minimum separation between approach/departure paths that are not
reciprocal in the FAA heliport and vertiport design
guidance \citep{faa2023heliport, faa2024eb105}; it is adopted here as a
directional separation constraint between candidates, not as a procedure
separation standard. The threshold keeps directions with a low integrated score
out of the operational set, and the separation criterion prevents two
candidates that effectively represent the same departure direction from being
selected together. No criterion yet prescribes a minimum integrated score, so
$\taumin$, the midpoint of the score range, was set for this study.

\begin{figure}[htbp]
	\centering
		\subfloat[Suseo station]{\includegraphics[width=0.49\linewidth]{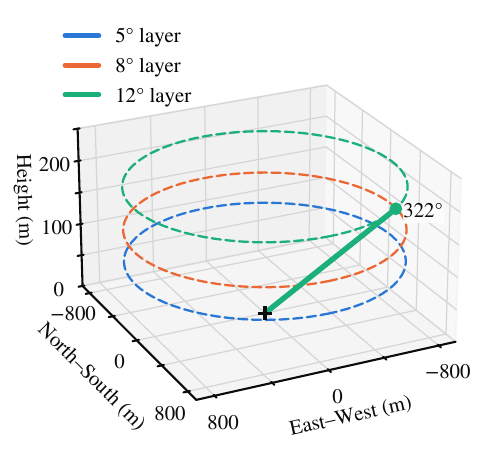}}\hfill
		\subfloat[Jamsil heliport]{\includegraphics[width=0.49\linewidth]{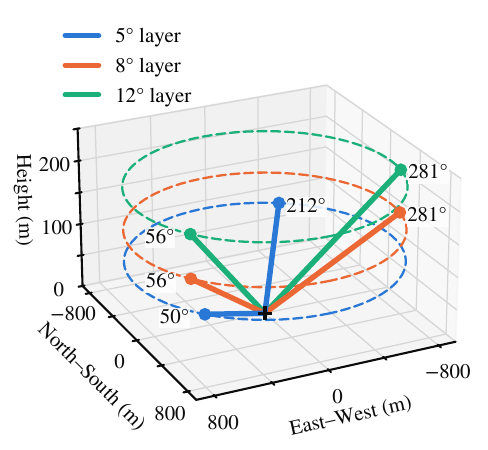}}
	\caption{Departure candidates selected under the obstacle prioritized scenario ($V_{cl} = 10.5\,\mathrm{kt}$).}
	\label{fig:normal}
\end{figure}

Table~\ref{tab:ranked} and Fig.~\ref{fig:normal} show the resulting
candidates. At Case 1 no azimuth reaches $\tau$ in the $5^{\circ}$ and
$8^{\circ}$ layers, whose highest integrated scores under this scenario are
0.308 and 0.473, so these layers yield no candidate. In the $12^{\circ}$ layer
24 azimuths reach $\tau$, all within $293$--$332^{\circ}$, so the layer yields a
single candidate, $322^{\circ}$. This direction lies close to the track of the
Tancheon route segment that passes the site in the aeronautical chart, about
$320^{\circ}$ from waypoint TC010 to TC009 \citep{molit2023uamchart}. At Case 2,
167, 204, and 227 azimuths reach $\tau$ in the $5^{\circ}$, $8^{\circ}$, and
$12^{\circ}$ layers, and the candidates are limited by the separation criterion
rather than by $\tau$: the two conditions select $50^{\circ}$ and $212^{\circ}$
in the $5^{\circ}$ layer and $281^{\circ}$ and $56^{\circ}$ in the $8^{\circ}$
and $12^{\circ}$ layers.

\begin{table}[htbp]
\caption{Departure candidates selected under the obstacle prioritized scenario
($V_{cl} = 10.5\,\mathrm{kt}$, $\tau = \taumin$, minimum separation
$135^{\circ}$). Integrated scores $S_{total}$ are given in parentheses.}
\label{tab:ranked}
\centering
\small
\begin{tabular}{cccc}
\toprule
Layer & Azimuths & First & Second \\
 & reaching $\tau$ & candidate & candidate \\
\midrule
\multicolumn{4}{@{}l}{\textit{Suseo station}} \\
$5^{\circ}$  & 0  & -- & -- \\
$8^{\circ}$  & 0  & -- & -- \\
$12^{\circ}$ & 24 & $322^{\circ}$ (0.567) & -- \\
\midrule
\multicolumn{4}{@{}l}{\textit{Jamsil heliport}} \\
$5^{\circ}$  & 167 & $50^{\circ}$ (0.841)  & $212^{\circ}$ (0.547) \\
$8^{\circ}$  & 204 & $281^{\circ}$ (0.842) & $56^{\circ}$ (0.836) \\
$12^{\circ}$ & 227 & $281^{\circ}$ (0.852) & $56^{\circ}$ (0.836) \\
\bottomrule
\end{tabular}
\end{table}

The two sites thus differ in the directional resource that remains after
selection. At Case 2 every layer keeps two separated candidates, and in the
$8^{\circ}$ and $12^{\circ}$ layers the second candidate scores almost as high
as the first (0.836 against 0.842 and 0.852). At Case 1 only the $12^{\circ}$
layer keeps a candidate, and the southern sector that the steeper layers reopen
(Sec.~\ref{subsec:obsresult}) does not reach $\tau$ because the low buildings
near the FATO leave it little obstacle margin.

The selection depends on the threshold $\tau$. Lowering $\tau$ to 0.3 under the
obstacle prioritized scenario admits $324^{\circ}$ in the $5^{\circ}$ layer and
$322^{\circ}$ in the $8^{\circ}$ layer at Case 1 and adds $187^{\circ}$ to
$322^{\circ}$ in the $12^{\circ}$ layer, whereas the candidates at Case 2 are
unchanged. The absence of candidates in the $5^{\circ}$ and $8^{\circ}$ layers
at Case 1 therefore reflects the chosen threshold as well as the obstacle
environment. The obstacle score formulation, the exponent $p$, and the weights
were set for this study, so the integrated score is a relative measure and the
value of $\tau$ appropriate to it requires further examination. At a site
constrained mainly by obstacles, such as Case 1, $\tau$ could be set below
$\taumin$, or the obstacle score could be adjusted, through its formulation or
through $p$, so that azimuths with a limited but secured margin remain in the
candidate set. Likewise, two candidates were sought in each layer here to
satisfy this $135^{\circ}$ separation, but where only one direction reaches
$\tau$, as in the $12^{\circ}$ layer at Case 1, designating that single
direction may be the practical choice for the site.

Finally, the candidates selected here result from a static evaluation of the
ten year wind record and the present building data. Because wind conditions and
airspace availability change over time in actual operations, an extension that
updates the candidate set with such variation could be addressed in future
work.

\section{Policy and Operational Implications}
\label{sec:implications}

\subsection{Facility Planning for Heterogeneous Fleets}
\label{subsec:imp_facility}

The directional resource of a vertiport depends on the climb capabilities it is
meant to accommodate as well as on its facility and surroundings. The same climb
profile secured very different shares of the azimuth space at the two sites, and
the weight at which the leading azimuth reverses differed across climb angle
layers, so a preferred departure direction may hold only for a limited set of
layers. Departure directions should therefore be designated together with the
range of vehicle types a facility will serve, and judged site by site; the
directional availability indicator and the candidate selection results of this
study can support parameter selection and policy decisions.

\subsection{Operational Response to Crosswind}
\label{subsec:imp_crosswind}

With a single fixed departure direction, a crosswind component approaching the
allowable limit leaves only a delay or a cancellation. A candidate set evaluated
in advance can instead be checked against the current wind at the operational
stage, and a usable direction with a smaller crosswind component chosen. The wind
analysis of this study serves long term planning: it presents, from the long
term record, the set of procedures available at a vertiport, and when real time
crosswind data become available, the procedure in that set best suited to the
wind of the moment is selected. Departing on such an alternative procedure reduces
ground holding, which keeps congestion from spreading through the network and
allows more operations.

\subsection{Higher Level Traffic Management and Airspace Integration}
\label{subsec:imp_atm}

Research on low altitude traffic management is developing toward an integrated
treatment of airspace sectorization, corridor capacity, and vertiport
constraints \citep{kim2026centralized}, and the performance of such models
depends on how many departure directions are usable at each vertiport and under
what conditions. The ranked candidate sets of this study can be retained as the
procedures available to such a model and embedded in an optimization, with the
choice of procedure as a decision variable and the usability of each procedure
as a constraint. From the operator's perspective,
the choice among these procedures can minimize eVTOL energy cost or maximize
operating profit: departing along the usable direction closest to the
destination reduces energy consumption; assigning departures
and arrivals to separated directions relaxes their timing constraints; and
several climb angle layers let an operator combine vehicle types within one
fleet. From the user's perspective the objective can be to minimize total
travel time, and at the network level to maximize UAM traffic volume or minimize
investment cost. Because every procedure in the candidate set already satisfies
the obstacle clearance requirement, these problems can be solved on a realistic
basis that combines demand and capacity with aeronautical factors.

Integrating UAM with the existing airspace also requires checking vertiport
departures for conflicts with conventional aviation and helicopter airspace.
Because the framework yields a ranked set of candidate procedures, any direction
that conflicts with such airspace can be removed and the next ranked directions
used in its place, so a variety of procedures can be designed and presented
flexibly. The framework evaluates every combination of azimuth, climb angle
layer, crosswind limit, and weighting scenario, yet the computational load is
small, so results are obtained quickly once the obstacle and wind data are
prepared. It can
therefore be run at many candidate vertiport sites as a preliminary validation
step before detailed procedure design.

\section{Conclusion}
\label{sec:conclusion}

This study addressed how operational flexibility in the departure direction can
be secured when eVTOL aircraft with different climb and crosswind capabilities
share a single vertiport. It proposed a framework that integrates urban obstacle
conditions and wind operability within a single scoring structure, evaluates the
full $360^{\circ}$ of azimuth for each climb angle layer, and derives a ranked
set of usable directions for each layer. Applied to two sites in Seoul, the
framework showed that the same climb performance can secure very different
directional resources at different sites and that no single departure direction
is preferred at a site; the choice follows from operational priorities, the
climb profile available to the aircraft, and the arrangement of obstacles and
winds around it.

This study is limited in that the climb angles and crosswind limits are
representative scenarios, building heights were simplified to $4\,\mathrm{m}$
per story without terrain elevation from a digital elevation model, and the wind
was taken, as a proxy, from the 16 point record of a synoptic station about
$11$--$15\,\mathrm{km}$ from the sites. Future work should incorporate type
specific performance as it emerges from certification and flight testing and
reflect the temporal variation of wind and airspace availability.

\section*{Funding Sources}

This work was supported in part by the National Research Foundation of Korea
grant funded by the Korea government (Ministry of Science and ICT)
(Grant No. RS-2024-00342879), and in part by the Korea Agency for Infrastructure
Technology Advancement grant funded by the Ministry of Land, Infrastructure and
Transport (Grant No. RS-2026-25527518).

\section*{Acknowledgments}

Aircraft symbols used in the figures were adapted from publicly available
materials from NASA's Revolutionary Vertical Lift Technology project. The
authors used Claude (Anthropic) to improve the readability of the manuscript;
the authors reviewed and edited the content as needed and take full
responsibility for the content of the publication.

\bibliography{refs}

\end{document}